\documentclass[apjl]{emulateapj}

\def\cygx1{Cyg X-1\ }

\def\mtav{}

\def\awc{}
\def\sasa{}
\def\sasasa{}
\def\sasasasa{}

\usepackage{ifpdf}

\usepackage{makeidx}

\usepackage{amssymb}

\usepackage{amsmath}

\usepackage{eurosym}

\usepackage{amsfonts}

\usepackage{rotating}

\usepackage{amssymb}

\newcommand{\flux}{$\rm ph$ $\rm cm^{-2}$ $\rm s^{-1}$ }
\newcommand{\mev}{$\rm MeV$}
\newcommand{\kev}{$\rm keV$}
\newcommand{\gev}{$\rm GeV$}
\newcommand{\ghz}{$\rm GHz$}
\newcommand{\hz}{$\rm Hz$}
\def\sasa{}
\def\mt{}
\def\mtt{}

\def\mta{}

\def\mtav{}

\shortauthors{S. Sabatini et al.}

\begin{document}



\title{\mt Gamma-ray observations of Cygnus X-1 above 100
{\bf \mev} in the hard and soft states}


\bigskip

\author{
S. Sabatini \altaffilmark{1,2}, M.~Tavani\altaffilmark{1,2,3}, P. Coppi\altaffilmark{4},
G. Pooley\altaffilmark{5}, M. Del Santo\altaffilmark{1}, R. Campana \altaffilmark{1,7},A. Chen\altaffilmark{6},
Y. Evangelista\altaffilmark{1},  G. Piano\altaffilmark{1,2},
 A. Bulgarelli\altaffilmark{7},
P. W. Cattaneo\altaffilmark{8}, S. Colafrancesco\altaffilmark{9},
E. Del Monte\altaffilmark{1}, A. Giuliani\altaffilmark{6}, M. Giusti\altaffilmark{1},
F. Longo\altaffilmark{10}, A. Morselli\altaffilmark{2,3}, A. Pellizzoni\altaffilmark{11},
M. Pilia\altaffilmark{12}, E. Striani\altaffilmark{1,3}, M. Trifoglio\altaffilmark{7},
S. Vercellone\altaffilmark{13}
}

\altaffiltext{1} {INAF/IAPS-Roma,I-00133 Roma, Italy}
\altaffiltext{2} {INFN Roma Tor Vergata, I-00133 Roma, Italy}
\altaffiltext{3} {Dip. di Fisica, Univ. Tor Vergata, I-00133 Roma,
Italy}
\altaffiltext{4} {Yale University, P.O. Box 208101, New Haven, CT 06520-8101, USA}
\altaffiltext{5} {Cavendish Laboratory, University of Cambridge,
Cambridge CB3 0HE, UK}
\altaffiltext{6} {INAF/IASF-Milano, I-20133 Milano, Italy}
\altaffiltext{7} {INAF/IASF-Bologna, I-40129 Bologna, Italy}
\altaffiltext{8} {INFN-Pavia, I-27100 Pavia, Italy}
\altaffiltext{9} {INAF-OAR, I-00040 Monteporzio Catone, Italy}
\altaffiltext{10} {Dip. Fisica and INFN Trieste, I-34127 Trieste,Italy}
\altaffiltext{11} {INAF-OAC, I-09012 Capoterra, Italy}
\altaffiltext{12} {ASTRON, the Netherlands Institute for Radio Astronomy, Postbus 2, 7990 AA, Dwingeloo, The Netherlands}
\altaffiltext{13} {INAF/IASF-Palermo, I-90146 Palermo, Italy}





\begin{abstract}

{\mtt We present the results of multi-year gamma-ray observations
by the {\it AGILE} satellite of the black hole binary system Cygnus X-1.
In a previous investigation we focused on gamma-ray
observations of Cygnus X-1 in the hard state during the period
mid-2007/2009.}
%
Here we present the {\mt results of the} 
gamma-ray monitoring of Cygnus X-1 during the period 2010/mid-2012 
{\mtt which includes a remarkably prolonged `soft state' phase
(June 2010 -- May 2011). Previous 1--10 \mev\ observations of Cyg X-1 in
this state hinted at a possible existence of a non-thermal
particle component with substantial modifications of the
Comptonized emission from the inner accretion disk. Our {\it AGILE}
data, averaged over the  mid-2010/mid-2011 soft state of Cygnus X-1,
provide a significant upper limit for gamma-ray emission
above 100 \mev\ of
$F _ {\rm soft} < 20 \times 10^{-8}$ \flux,
excluding
the existence of prominent non-thermal emission above 100 \mev\
during the soft state of Cygnus X-1. We discuss theoretical
implications of our findings in the context of high-energy
emission models of black hole accretion.}{\mtt We also discuss 
possible gamma-ray flares detected by {\it AGILE}.
In addition to a previously reported episode observed by {\it AGILE} in October 2009 during the
hard state, we report a {\mta weak but important candidate
for} enhanced emission which occurred at the end of June 2010 (2010-06-30 10:00 -
2010-07-02 10:00 UT) {\mta exactly in coincidence with a}
hard-to-soft state transition and before {\mta an anomalous}
radio flare. An appendix summarizes all previous high-energy observations and
{\mta possible} detections of Cygnus X-1 above 1 \mev .}

\end{abstract}


\section{Introduction}
Cygnus X-1 (Cyg X-1) is the archetypal black hole binary system in our Galaxy. It is composed
of a compact object and a O9.7 Iab supergiant star companion with a mass estimate ranging
 between $\sim 17-31$  $\rm M_{\odot}$, filling 97$\%$ of its Roche Lobe (Gierlinski et al., 1999; 
Caballero-Nieves
et al., 2009). The measurements of the mass for the compact object
range from 4.8 to 14.8 $M_{\odot}$ (Herrero et al., 1995;
Shaposhnikov \& Titarchuk, 2007; Orosz et al., 2011), suggesting identification with a
black hole. Being one of the brightest sources in the X-ray sky and
having a persistent emission, the literature on the system is
extremely rich and extensive monitoring in radio, IR, UV and X-rays has been carried out
 (Mirabel et al., 1996; Pooley et al., 1999; Fender et al., 2000; {\mtt McConnell} et al., 2002;
Gallo et al., 2003; Pandey et al., 2006; Del Monte et al., 2010; Rahoui et al., 2011; 
 Jourdain et al., 2012), leading to interesting
correlations and being of great importance for understanding the
 process of accretion onto black holes in general.



Typical {\mtt X-ray} spectral states of Cyg X-1 have been
classified into the `hard/low' and
`soft/high' states, which are defined according to the spectral
behaviour at X-ray energies ($< 20$ \kev). The source usually
spends 90$\%$ of its time in the low/hard spectral state {\sasa
whose spectral energy distribution is} well described by a power-law (E$^{-\gamma}$) with
photon index ${\gamma} \sim$1.7, {\mtt a very prominent {\sasa broad} emission
peak of the power spectral energy density ($\nu \, F_{\nu}$)
near 100 \kev}, and a high-energy cutoff at
 $\sim$150 \kev. The less common soft state is characterized by {\mtt the absence of
 the prominent peak near 100 \kev,} a strong
 blackbody component with kT $\sim$ 0.5 \kev, and a soft power-law tail with $\gamma$ usually ranging
between 2 and 3. Intermediate spectral states also exist (see,
e.g., Belloni et al., 1996) {\sasasa and a number of different
spectral shapes have been reported in the literature {\mtav (e.g.,
INTEGRAL} observations, Del Santo et al., 2013 and references
therein).}

The different {\mtt spectral} states are usually described by the
interplay of a {\mtt relatively cool  accretion disk and a hot
optically
thick corona  surrounding the central source}. In the hard state,
the spectral energy distribution can be modeled by Comptonization
of abundant {\mtt soft} blackbody photons from {\mtt the inner}
accretion disk {\sasa which scatter off the} energetic electrons of
{\mtt the optically thick corona (e.g., Coppi 1999, 2006; Zdziarski et
al. 2002, 2004, 2011, 2012). A crucial property of this corona, energized by the
accretion process onto the black hole, is its ability to add a
non-thermal tail to an otherwise thermal distribution of electrons,
possibly extending to the gamma-ray energy range. This process of non-thermal
energization of coronal electrons is strongly constrained in the
Cyg X-1  hard states by the apparent cutoff observed above 150 \kev\
(Gierlinski et al., 1997; McConnell et al., 2002)  and by the absence of detectable
gamma-ray emission above 100 \mev\ (Sabatini et al., 2010a)}. In the transition to the soft
state, {\mtt the Comptonizing corona 
shrinks, } 
the cool disk moves inwards {\mtt (}possibly very
close to the last stable orbit{\mtt), and
non-thermal processes, if existing, can be revealed by emission above the disc 
blackbody component, in particular with the
detection of prominent power-law components above the \mev\ energy range in the
soft spectral state.}

{\mtt For many years, {\sasa the only available} information on}  the spectral states of
Cyg X-1 {\mtt above \mev\ energies}  
was
 the data collected {\mtt by the
COMPTEL instrument on board of the {\it CGRO}} (Collmar,
2003). 
{\mtt Cyg X-1 remained in the hard state for most of the {\it CGRO}
observations, as monitored by the hard X-ray instrument BATSE
(McConnell et al., 2002). However, during the {\it CGRO} lifetime,
two transitions to Cyg X-1  soft states were
studied by the combined effort of the OSSE, COMPTEL and EGRET
instruments (see the Appendix for more details of these important
observations).
Cyg X-1 transitions to the soft state are relatively rare (e.g.,
Zhang et al., 1997a), and not well understood theoretically. A very
significant non-thermal emission episode was detected by COMPTEL
in one case\footnote{ \mtt In the following, we are going to take
the COMPTEL detection of Cyg X-1 in the soft state reported by
McConnell et al. (2002) as a typical soft-state emission by a
non-thermal component.} with a maximum photon energy recorded
at 5--10 \mev\ (McConnell et al.1997; 2002).
This detection was for many years the only indication of a
possible non-thermal component in the soft state spectrum of
Cyg X-1, and stimulated many investigations and speculations about
its nature (Gierlinski et al., 1999; 
Zdziarski et al., 2002).
In particular, {\awc the detection of emission up to 100 \mev\ and beyond
would test}
hybrid Comptonization spectral models of black hole emission.
{\awc As a result, there has been great} interest in
 new gamma-ray data from Cyg X-1 in a soft state by the current
generation of gamma-ray space instruments ({\it AGILE} and {\it Fermi}).

In a previous paper we reported on the
gamma-ray observations of Cyg X-1 by the {\it AGILE} satellite that were
obtained during the period 2007--2009, during which the source was
in a prolonged hard state (Sabatini et al., 2010a). Here we present
the results of the {\it AGILE} gamma-ray monitoring of
Cyg X-1 during the 2010/mid-2012 period.
}
{\awc This period includes the June 2010 event during which}
the system underwent a clear spectral transition from the hard to
the soft state and unusually remained in the soft state for
almost a year. {\mtt This gave us the unprecedented opportunity to
carry out a long term monitoring of the soft spectral state of
Cyg X-1  at gamma-ray energies and
investigate on the possible existence of prominent emission above
100 \mev .}


Gamma-ray data in the Cyg X-1 soft state are of crucial
importance for theoretical modeling because they constrain
the high energy part of the spectrum, most likely dominated by
non-thermal emission. Of particular interest are observations that
can determine a clear cutoff in the spectra at high energies,
since the cutoff energy is a function of the compactness of the
inner source region.

{\mtt For a proper evaluation of the physical properties of Cyg X-1
in different accretion states, it is important to consider also
radio and X-ray emission in addition to gamma-ray data above 50
\mev . In particular,} for many years Cyg X-1 has been monitored in
search of non-thermal radio jets. Radio emission is observed to be
persistent with a modulation related to the orbital period of the system 
(Zhang et al., 1997b; Stirling et al., 2001) 
during the hard states and presents a strong decrease during soft states  
(see e.g., Zdziarski et al., 2011). {\sasasasa {\mtt Definitive evidence for 
a resolved extendend relativistic radio jet was provided by} Stirling et al. 
(2001) using VLBA and MERLIN data.  }
Fender et al. (2001) {\sasasa estimated an angle of 30$^{\circ}$
{\mtt between the jet axis and} the line of sight, assuming
the jet to be perpendicular to the disk. A more recent estimate for the angle of inclination of
the orbital plane to our line of sight is $27.1 \pm 0.8^{\circ}$ (Orosz et al., 2011).}
{\mtt A jet bulk Lorentz
factor of $\Gamma = (1- \beta^2)^{-1/2}\simeq 1.25$, and a jet
kinetic power  $P_{\rm j} \simeq (1-3) \times 10^{37} \, \rm erg \, s^{-1}$ 
have been determined in the hard state
from the large scale
optical emission of a nebula most likely energized by the Cyg X-1 jet
(Gallo et al., 2005; Russell et al., 2007; see also Gleissner et al.,
2004; Malzac et al. 2009; and the discussion in Zdziarski et al.
2012).}

{\mtt Cyg X-1 has been repeatedly observed in X-rays both in the hard
and in soft states. Of particular
interest are the {\it INTEGRAL} observations of Cyg X-1  that cover the
energy range 20 \kev\ -- 1 \mev\ (see the recent review and discussion
by Zdziarski et al. (2012) who also reconsider the spectral data of
Laurent et al., 2011).}  {\awc An important aspect of high-energy
emission from Cyg X-1 is its variability.} Variability in the X-ray band has been
observed on several different timescales (Brocksopp et al., 1999;
Pottschimidt et al., 2003; Ling et al., 1997; Golenetskii et al., 2003).
{\sasasa {\mtt Several} outburst episodes {\awc in both the
hard and soft states at various orbital phases were also}
reported by Golenetskii et al. (2003) using the
Interplanetry Network in the 15--300 \kev\ band and by Gierlinski \& Zdziarski (2003)
in the {\it RXTE/}PCA 3--30 \kev\ data.} {\mtt Variability of the
high-energy emission from Cyg X-1 is indeed a crucial issue.} More
recently very fast transient activity (on the order of hours) was
also detected at the TeV energy range by the MAGIC telescope
(Albert et al., 2007), and in the radio frequency by the MERLIN and  Ryle telescopes
(Fender et al., 2006).

\begin{figure*}[t!]
\begin{center}
\includegraphics [width=13 cm] {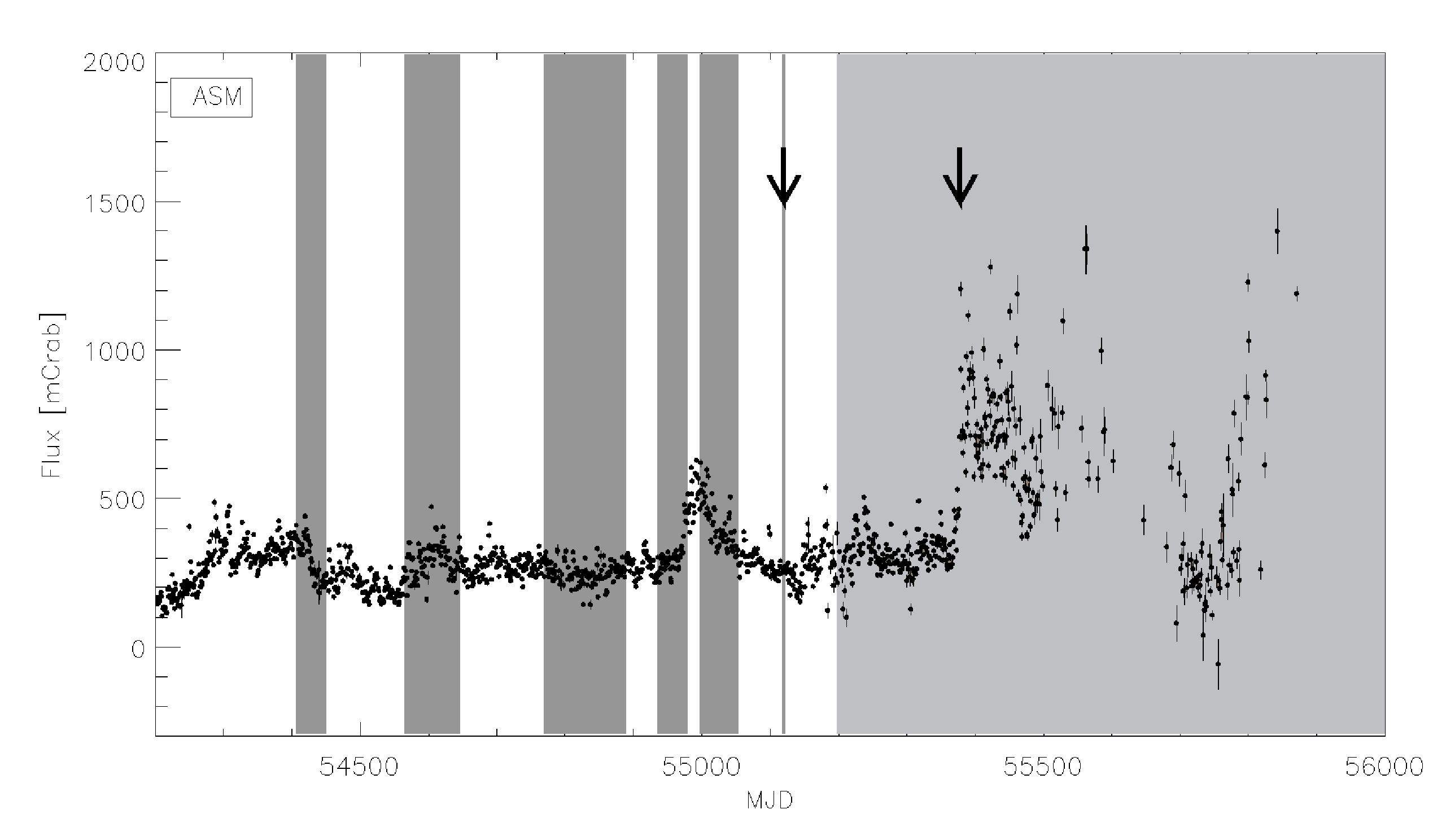}
\includegraphics [width=13 cm] {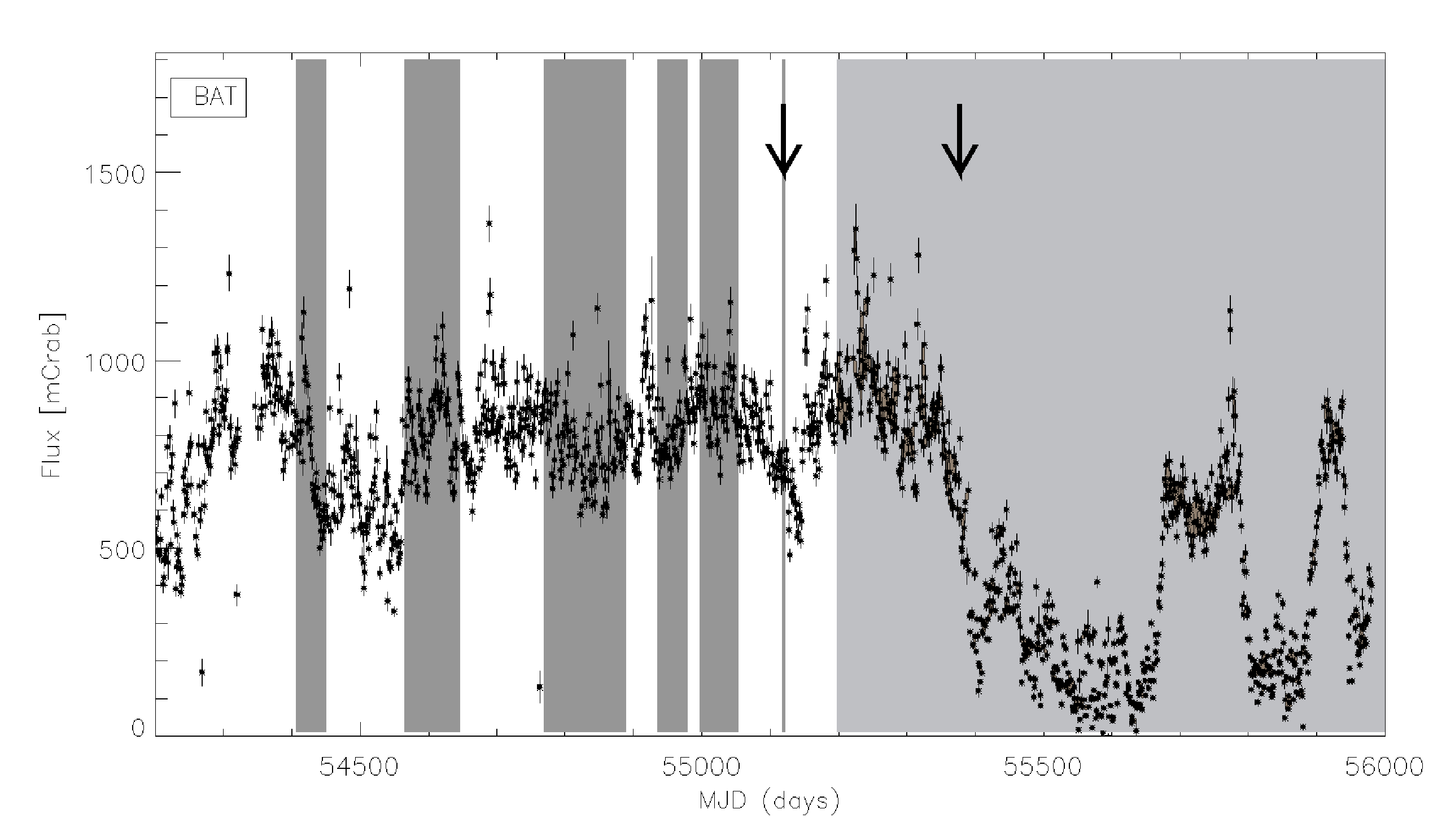}
\caption{Long term daily monitoring of Cyg X-1 in the soft and
hard X-ray bands. Upper panel shows {\it RXTE}-ASM data in the 1.3-12.2
\kev\ energy range; lower panel shows {\it Swift}-BAT data in the 15-50
\kev\ energy range. The grey areas show {\it AGILE} observing intervals
covering the Cygnus region: dark grey regions refer to pointing
mode and light grey to spinning mode of the satellite
respectively. Black arrows show the gamma-ray flares observed by
{\it AGILE} {\mtt as} reported in this paper.} \label{ASM_BATlongterm}
\end{center}
\end{figure*}

{\mtt For a black hole mass $M \sim 10 \, \rm M_{\odot}$, both the
total X-ray emission $L_{\rm X} \simeq 10^{37} \, \rm erg \, s^{-1}$ and
jet kinetic power in the hard state $P_j$ indicate sub-Eddington
accretion conditions. Data in the soft state of Cyg X-1 show that the
X-ray luminosity can be similar or typically higher and a low-level jet
activity can be present during this radio quenched state (Rushton et al. 2011; 2012;
see also below and the Appendix, 
sec \ref{sec:softstatedata}). In general, we can distinguish two types 
of gamma-ray emission from a black
hole system such as Cyg X-1: \textit{(1)} `accretion-driven
emission', with X-rays and possibly gamma-rays originating from
the inner accretion disk and/or Comptonizing corona; \textit{(2)}
`jet emission' originating in the accelerating flow of the
jet\footnote{Interaction of a non-thermal
relativistic jet with the ambient photon fields from the accretion
disk, the corona and the companion star wind contributing to the
high energy band of the spectrum (hard X-rays \/ $\gamma$-rays),
can be modeled both in hadronic (Romero et al. 2003, Perucho \&
Boch-Ramon 2008) or leptonic scenarios (Perucho \& Bosch-Ramon
2008; Piano et al., 2012; Zdziarski et al., 2012; Zdziarski,
2012).}.
%
{\awc The interpretation of the 1--10 \mev\ emission and above plays a
crucial role}.
This spectral component, detected both in the
hard and in the soft states of Cyg X-1 (see below) can be attributed
to hybrid Comptonization of accretion-driven emission or to a
synchrotron tail of jet emission (e.g. Zdziarski et al., 2012).
In this paper we focus especially on the gamma-ray emission of the
Cyg X-1 soft state during which jet activity is in general
subdued compared to the hard state (see e.g. Fender et al., 2004). We
therefore aim here at  constraining the possible existence of an
accelerated population of electrons/positrons for the
accretion-driven scenario.}

{\mtt Section 2 reviews the {\it AGILE} gamma-ray observations of Cyg X-1
in the hard state as well as during the recent prolonged (almost
1-year long) soft state period. We present in Section 3  the
theoretical implications of our upper limits to the emission above
100 \mev . Section 4 presents a general discussion of the
accretion-driven high-energy emission from Cyg X-1. We find it useful
to summarize all relevant previous gamma-ray
observations and detections of Cyg X-1 above 1 \mev\ in the Appendix. We also present
there two transient episodes of gamma-ray emission from
Cyg X-1, that at the moment constitute noticeable exceptions to the
standard low-intensity gamma-ray state. In particular, we present
data on a new relatively low-intensity/low-significance episode of
emission that occurred just prior to a major X-ray and radio
flaring transition on June 30 to July 2 2010.}


\section{{\it \bf AGILE} observations and results}

The {\it AGILE} {\mtt gamma-ray astrophysics} mission has been operating
since 2007 April (Tavani et al., 2008). The {\it AGILE} scientific instrument is very compact and is
characterized by two co-aligned imaging detectors operating in the
energy ranges 30 \mev\ -- 30 \gev\ (the imaging gamma-ray detector -
{\it GRID}; Barbiellini et al., 2002; Prest et al., 2003; Bulgarelli et al., 2010) and 18--60 \kev\
(the hard X-ray detector {\it Super-AGILE}; Feroci et al., 2007). {\mtt
An anticoincidence system (Perotti et al., 2006) and a
calorimeter sensitive in the 0.4--100 \mev\ energy range
(Labanti et al., 2006) complete the instrument}. {\it AGILE}'s performance is
characterized by large fields of view (2.5 and 1 sr for the
gamma-ray and hard X-ray bands, respectively), {\mtt good
sensitivity in pointing mode\footnote{\mtt {\it AGILE} operated in
pointing mode during the first phase of operations (July 2007 --
mid Oct 2009).
 Since January 2010 the satellite has been
operating in `spinning' mode, observing a large fraction of the
sky continuously with somewhat reduced sensitivity per unit time
but much increased overall sky coverage.} near 100 \mev\ (the on-axis
effective area is about 400 cm$^2$ at 100 \mev), and
state-of-the-art angular resolution} (68$\%$ containment radius PSF$\sim 3.5$ deg
at 100 \mev\ and PSF$\sim 1.5$ deg at 400 \mev).

 Flux sensitivity for a typical 1-week
observation in pointing mode can reach the level of $F  \sim
(20-30) \times 10^{-8}$ \flux above 100 \mev\
depending on off-axis angles and {\mtav pointing directions} (see
Tavani et al. (2008) for details about the mission and main
instrument performance).

\begin{figure*}[t!]
\begin{center}

\includegraphics [width=12.cm]{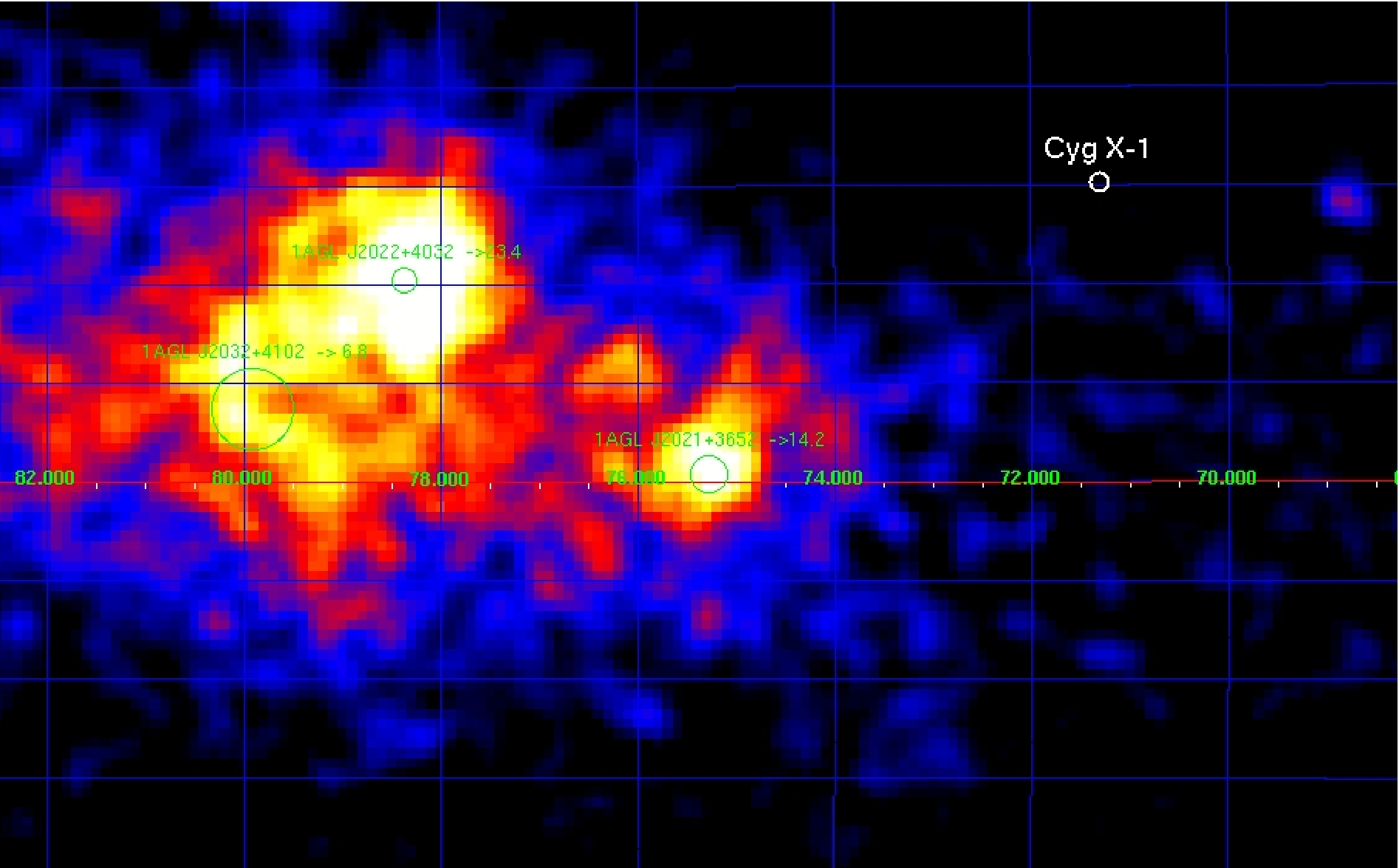}
\hspace{0.8cm}
\includegraphics [width=12.cm]{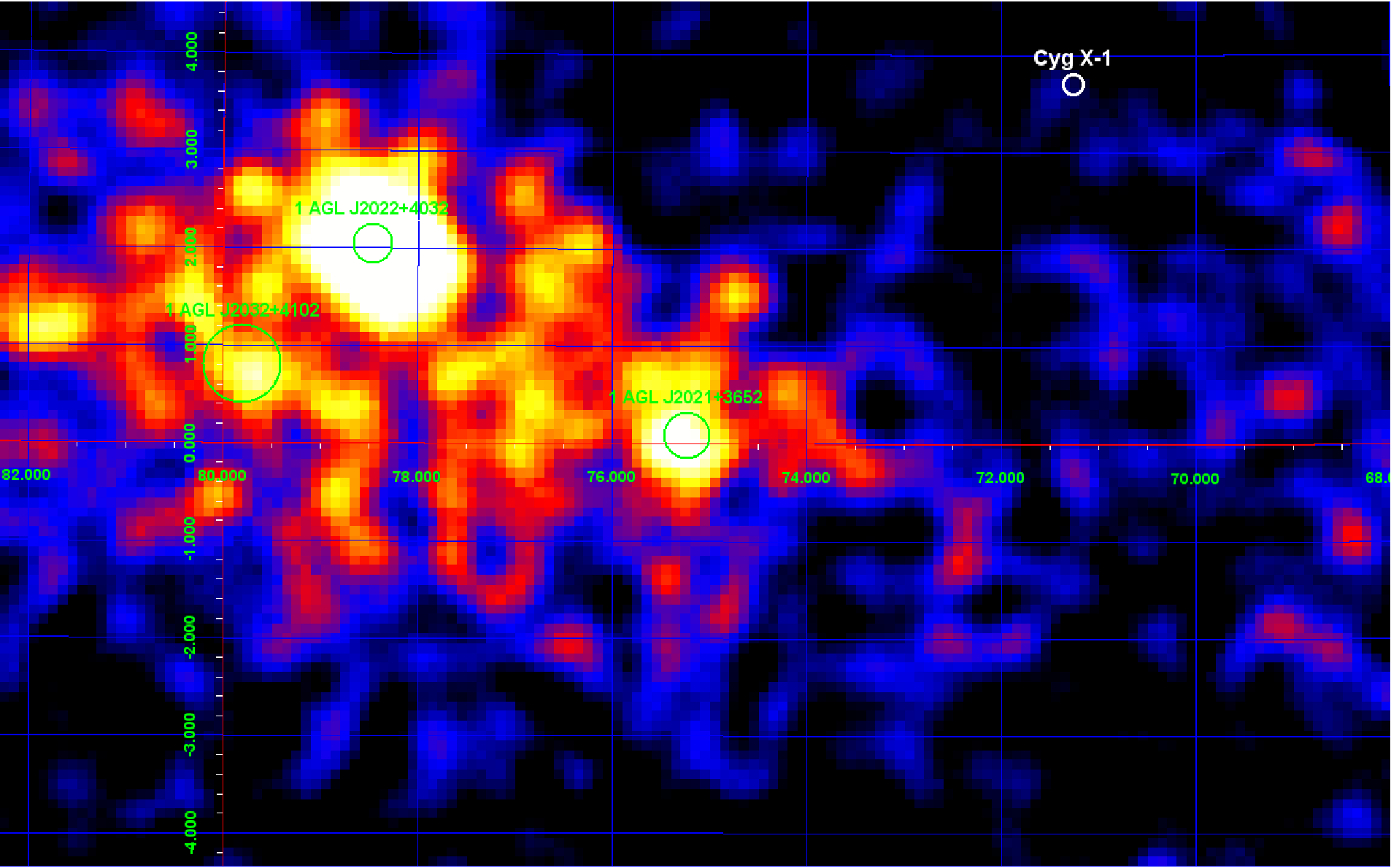}
\caption{{\it AGILE} gamma-ray deep intensity maps above 100 \mev\ of the
Cygnus Region in Galactic coordinates displayed with a three-bin
Gaussian smoothing. Pixel size is 0.1$^\circ$ and the nominal
position of Cyg X-1 is {\mtt marked} in white.
{\it Upper panel:} an integration of {\it AGILE} data covering all the
data of the pointing mode (2007--2009), when Cyg X-1 was  in the
hard state. {\it Lower panel:}  deep integration of {\it AGILE} data in
spinning mode selecting the time intervals during which
Cyg X-1 was in the soft state (MJD 55378--55647, see Fig. 1).
 }
\label{mappadeep}
\end{center}
\end{figure*}

{\it AGILE} 
observed the Cygnus region in the Galactic plane
several times
during the period 2007 July -- 2011 May (Sabatini et al., 2010a;
Chen et al., 2011; Piano et al., 2012). {\mt Fig \ref{ASM_BATlongterm} shows the
daily monitoring in the soft (ASM 1.3--12.2 \kev) and hard
({\it Swift}-BAT 15--50 \kev) X-ray {\mtt range}.  {\mtt {\it AGILE} observation intervals
of the Cygnus region in pointing (dark gray) and spinning (light gray) mode
are shown.}
The transition to (and
persistence in) the soft state starting around MJD 55380 is {\mtt
evident}.
In the first paper (Sabatini et al., 2010) we analyzed our pointing
mode data {\mtt up to the end of 2009 (MJD 55120)}.
Here we focus on the 2010 Jun -- 2011 May period, {\mtt during
which Cyg X-1 was entirely in the soft state}.

The analysis of the gamma-ray data presented in this paper was carried
out with the {\mtt standard} {\it AGILE-GRID} FM3.119 filter$\_$I0010 B20
calibrated filter with a gamma-ray event selection that takes into
account South Atlantic Anomaly event cuts and 80 degree Earth
albedo filtering. Throughout the paper, statistical significance
and source flux were determinated using the
standard {\it AGILE} multi-source likelihood analysis software
(Bulgarelli et al., 2012a). The
statistical significance is expressed in terms of a Test Statistic (TS)
(Mattox et al., 1996) and asymptotically distributed as a
$\chi^2/2$ for 3 degrees of freedom ($\chi_3^2/2$).
{\awc We assessed the pre- and post-trial significance using
multiple Monte Carlo simulations of the sensitivity of the gamma-ray
instrument to point-like source emission (Bulgarelli et al., 2012).}

\begin{figure*} 
\begin{center}
\includegraphics [width=12 cm, height=9 cm]{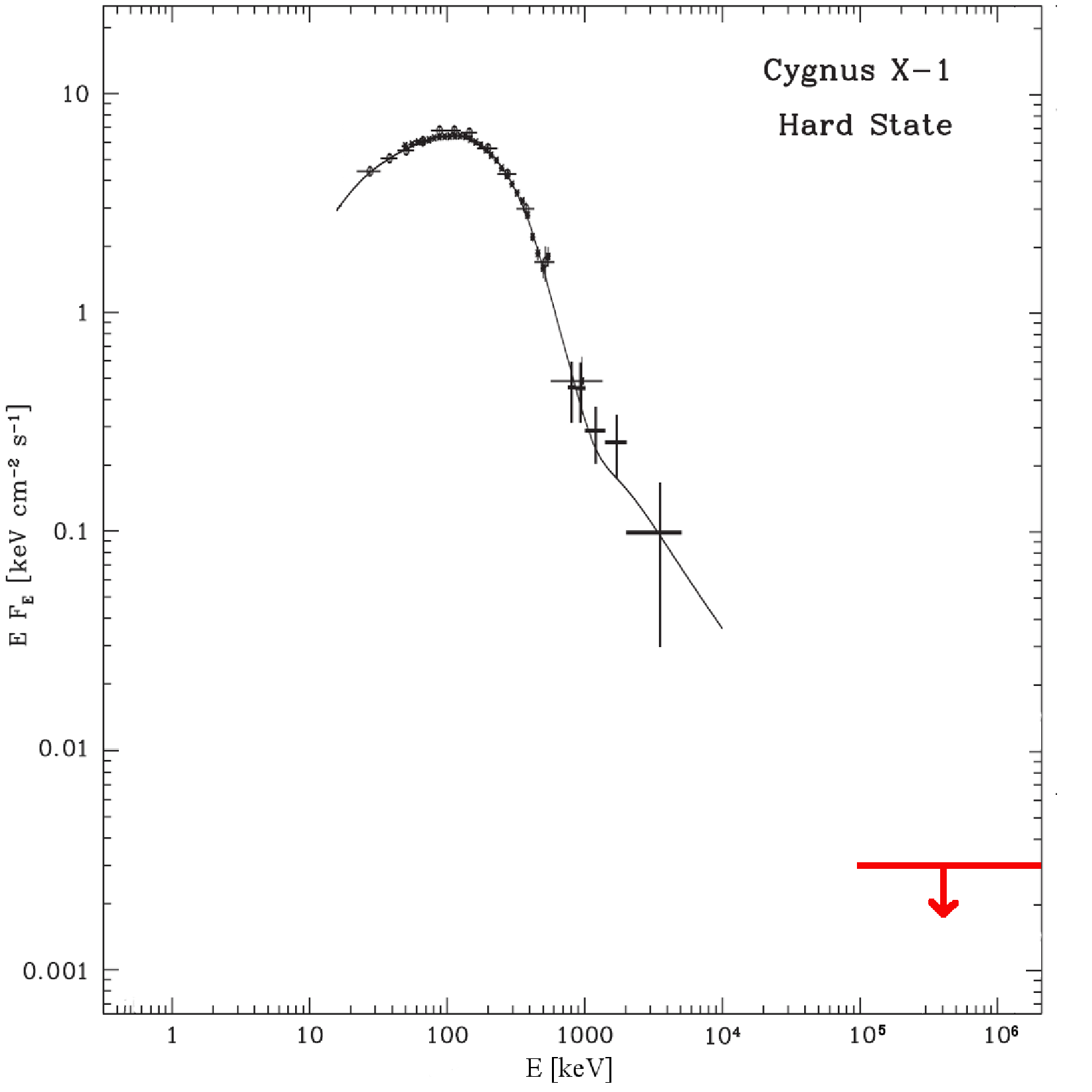}
\includegraphics [width=12 cm, height=9 cm]{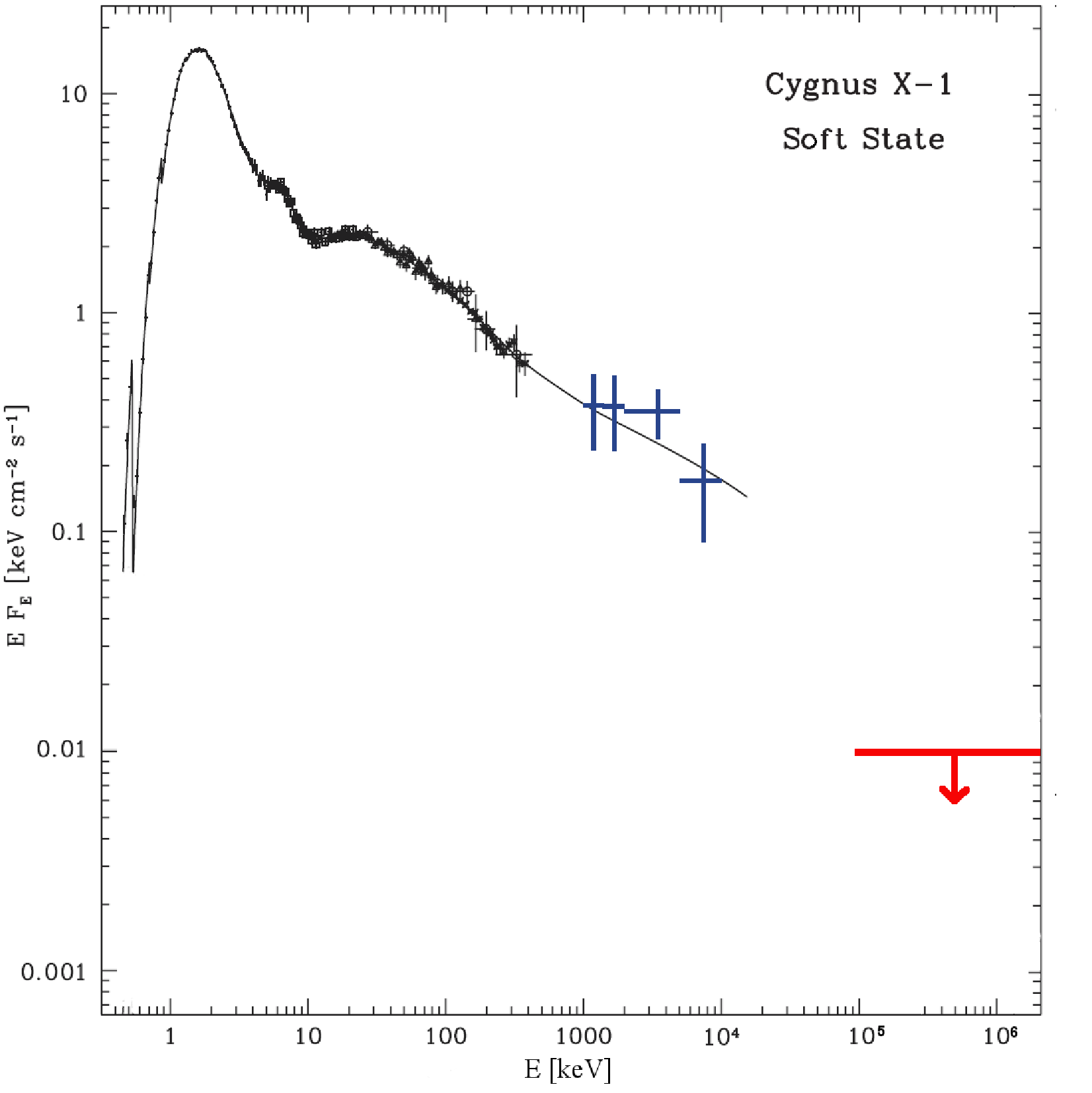}
\caption{ \mtt
Spectral energy distributions of Cyg X-1 for the hard and soft states with
superimposed {\it AGILE} upper limits (in red color). Solid lines are from McConnell et al. (2002).
{\it Upper panel:} data for the hard state include OSSE and COMPTEL (COMPTEL data
for this case are the average of nine different {\it CGRO} observations);
{\it Lower panel:} data for the soft state, including {\it LECS},
{\it HPGSPC} and {\it PDS} instruments on board {\it BeppoSAX} and OSSE,BATSE and COMPTEL
instruments on {\it CGRO} (data are for the soft state event detected
in June 1996).}
\label{McConnell}
\end{center}
\end{figure*}


Fig. \ref{mappadeep} shows {\mtt the {\it AGILE}} deep gamma-ray
integrations of the Cygnus region above 100 \mev\ during the periods
2007 July -- 2010 Oct (MJD: 54406 -- 55121) and 2010 June -- 2011 May
(MJD: 55378 -- 55647), covering  the hard and the soft spectral state
respectively. No gamma-ray persistent emission from Cyg X-1 was
detected by {\it AGILE} during either spectral states of the source for
these deep integrations. A multi-source likelihood analysis,
 including all known gamma-ray sources of the region,
provides a 2$\sigma$ upper limit for the  energy $\ge$ 100 \mev\
of $F_{\rm UL,hard} = 3 \times 10^{-8}$ \flux for the hard state
(Sabatini et al., 2010a) and
 $F_{\rm UL,soft}= 20 \times 10^{-8}$ \flux for the soft state.
Fig. \ref{McConnell} shows  typical hard and soft spectral states
from the literature (e.g., McConnell et al., 2002) together with
the {\it AGILE} upper limits (plotted in red). For the soft state,
we also plot in Fig. \ref{McConnell} (bottom panel) the soft
gamma-ray emission detected on one occasion by COMPTEL (McConnell et al.,
2002; see also the discussion in the Appendix).

The {\mtt {\it AGILE}} gamma-ray upper limit in the soft state is quite
important, and excludes a simple power-law extrapolation of the
soft gamma-ray emission detected by COMPTEL. Both measurements,
obtained with {\it AGILE} data after many months of observations,
confirm that Cyg X-1 is not a steady gamma-ray emitter above 100
\mev\ at levels comparable to those detected from {\mtt the other
prominent micro-quasar} Cygnus X-3 (Tavani et al., 2009; Abdo et
al., 2009; Bulgarelli et al., 2012b; Corbel et al., 2012;
Piano et al., 2012) . These findings have important theoretical
implications that we discuss in the next section.







\begin{figure*}[t!]
\begin{center}
\includegraphics [width=15 cm]{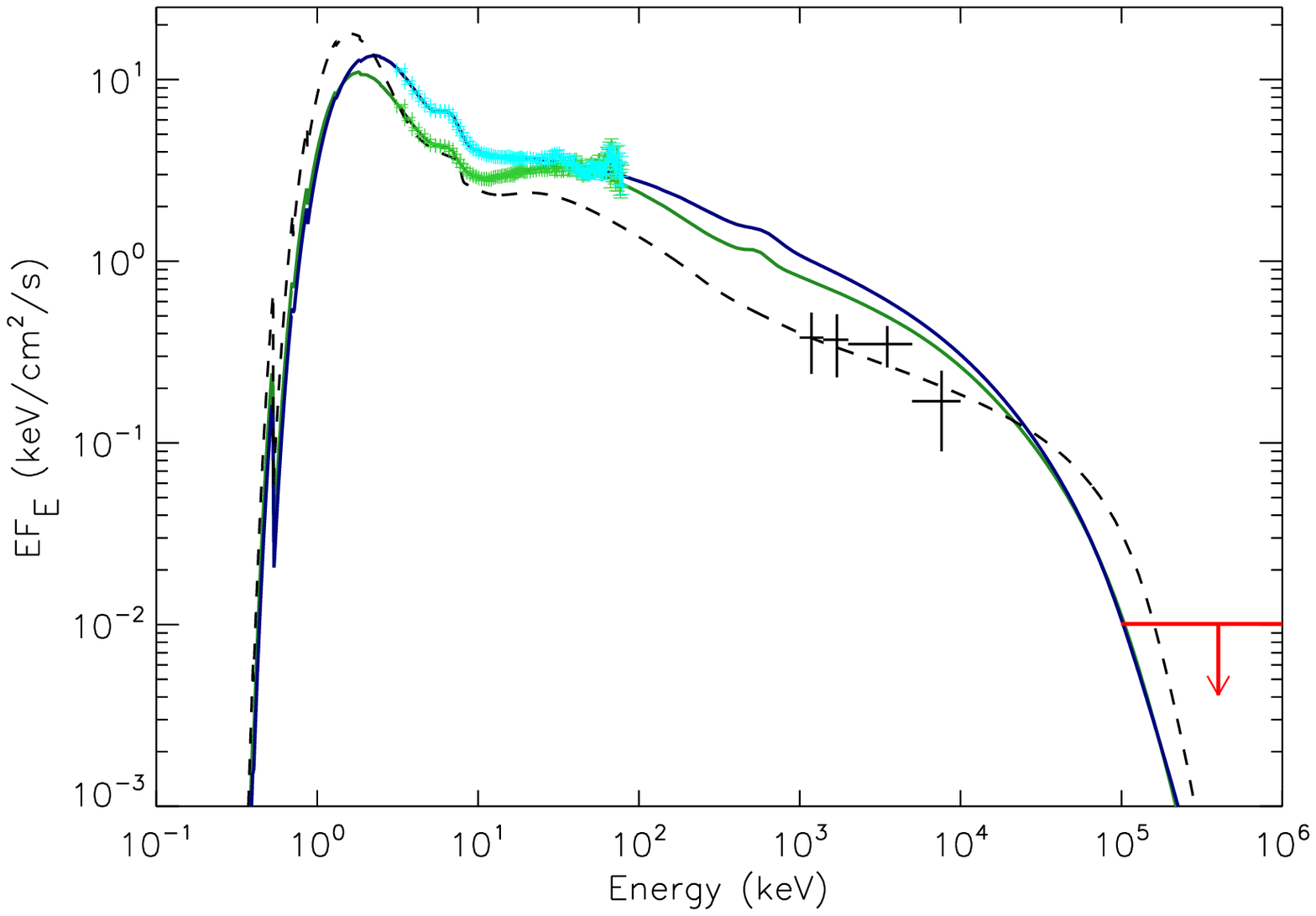}
\caption{{\sasasa The {\it AGILE} gamma-ray upper limit in the context 
of Cyg X-1 soft state data and modeling.
{\it RXTE} PCA/HEXTE data during the{\it AGILE} monitoring are for 
the 4$\rm ^{th}$ and 22$\rm ^{nd}$ of July
2010, in green and cyan respectively.
The solid line spectra are obtained with {\mtt  EQPAIR} with the parameters
of model-1 and -2, as discussed in the text}.
{\mtt X-ray absorption is taken into account} in this 
{\mtt calculation}.
As a comparison we also show the canonical soft state spectrum
(McConnell et al., 2002) with a dashed line and COMPTEL data in black.
}
\label{fig:gierlinski02}
\end{center}
\end{figure*}

\section{ {\it \bf RXTE} PCA/HEXTE data}
{\sasasa
Nineteen pointed observations were performed by {\it RXTE} PCA/HEXTE during
the period 2010 June 19 -- 2010 July 31, for a net exposure time of
about 68.5 ks, catching the source across the whole transition from the
hard to the soft state. The change of state can be described by a change in the 
Power Density Spectra (PDS) as shown in Fig. \ref{fig:PCA} in the Appendix 
and here we adopt Shaposhnikov \& Titarchuk (2006) nomenclature for the classification
of spectral states.
 The fractional RMS dropped to about 4\% on 2010 July 4, which clearly shows that the source
\mt{had} finally reached the soft state. Fig.\ref{fig:gierlinski02} shows {\it RXTE} PCA/HEXTE 
data of the 4$\rm th$ and 22$\rm nd$ of July, when the source was respectively in the soft 
and super-soft state
, during the AGILE monitoring .
}

\section{Results and Discussion}


{\mtt The lack of detectable gamma-ray flux above 100 \mev\ from
Cyg X-1 in the soft state leads to important theoretical
constraints.}
Cyg X-1 has been
{\mtt considered as a crucial test case}
for the modeling of radiation mechanisms {\mtt of accreting}
 black holes in the literature
 (Coppi et al., 1999; Gierlinski et al., 1999; Zdziarski et al., 2012
and ref. therein). {From the properties of the soft X-ray and hard X-ray
emission and {\mtt the  well defined pattern of} spectral state changes,
Comptonization models have been {\mtt successfully} applied to describe
the high-energy emission from Cyg X-1 (e.g., Coppi et al., 1999; Poutanen
et al., 1998; Zdziarski et al. 2002, 2012).
In this approach, different spectral states of the source are
interpreted in relation to the interplay between the emission from
an optically thick, cold accretion disc and a geometrically
thin/optically thick corona above the disc. {\mtt In
the simplest versions of this model,} the high energy emission of
the soft state is {\mtt expected to be} steady and possibly to
extend up to gamma-ray energies {\mtt above 1 \mev\ depending on
the details of the thermal vs. non-thermal electron/positron
component in the Comptonized corona}. The disk contributes
typically to the soft photon emission with a thermal distribution of
temperature $T_s$ and luminosity $L_s$. The corona is a much more
complex and dynamical system where non-thermal particle
acceleration, electron/positron pair formation and annihilation,
optically thick Comptonization of thickness $\tau$, and inverse
Compton scattering occur. It is customary to define a `hard
luminosity' $L_h$ that takes into account the emission originating
from these processes. Comptonization modeling using EQPAIR
numerical code (Coppi 1999) treats self-consistently these
processes, and can be used for the interpretation of Cyg X-1
observations. The system `compactness parameter' $l$ defined as $l = L
\sigma_T / R m_e c^3$ plays a crucial role, where $L$ is the
luminosity of interest (`soft' or `hard'), $\sigma_T$ is the
Thomson cross section, $R$ is the typical radius of interest (either the inner disk
and/or the corona) $m_e$ the electron's mass, and $c$ the
speed of light. Depending on the choice of {\mtt $L_s$ or $L_h$}
(and in principle of the corresponding emitting radius $R$) we can
define the `soft' ($l_s$) and `hard' ($l_h$) compactness
parameters. Constraining these values for the typical emission of
Cyg X-1 is a long-standing theoretical problem.}

 \begin{table*}[t!]
\begin{center}
\small
 \small
{\bf TABLE 1}\\
\begin{tabular}{|c|c|c|c|c|c|c|c|}
 \hline
&  $kT_s (keV)$ & $l_s$& $l_{h}/l_{s}$ &  $l_{nth}/l_{h}$  & $\Gamma_{inj} $ &
$\tau_{i}$& $\Omega/2\pi$\\[0.5ex]
 \hline
model-1 &   0.43$^{+0.01}_{-0.04}$&  (10) &0.56$^{+0.04}_{-0.07}$  & (0.99) & (2.7) & $0.85 \pm 0.20$  & 0.6$ \pm 0.1 $   \\
model-2 & 0.65$\pm 0.09$    &  (10)   &   0.57$^{+0.03}_{-0.05}$   & (0.99)     &  (2.7)    &  $<$0.3     &  0.3$\pm 0.1$     \\
model-3 &0.37 & 3.2 & 0.17 & 0.68 & 2.6   & 0.11  &1.3  \\

\hline
\end{tabular} \label{tab:tabmodelli}
\end{center}
\vspace{0.1cm}\small{
Comptonization Model Parameters (EQPAIR) for the soft spectral states shown in Fig.
\ref{fig:gierlinski02} and  Fig. \ref{fig:PCA}. Parameters among brakets are frozen in the fit; 
free paramenter errors are given at the 90$\%$ confidence level.
$KT_s$: disc blackbody
temperature; $l_s$: soft photon compactness; $l_h/l_s$: ratio of hard-to-soft
compactness; $l_{nth}/l_{h}$: ratio of non-thermal-to-total hard compactness;  $\Gamma_{inj}$:
injection index of electron power-law distribution;
$\tau_{i}$: optical depth; $\Omega/2\pi$: Compton reflection.
Model-1 refers to a fit to the {\it RXTE} PCA/HEXTE data of the soft state of the 4th of July 2010
(green solid line in Fig. \ref{fig:gierlinski02}) ;
model-2 is for the super-soft state of the 22nd of July 2010
(blu solid line in Fig. \ref{fig:gierlinski02}); model-3 reports McConnell et al. 
(2002) parameters as a reference (black dashed line in Fig. \ref{fig:gierlinski02}).\\
\vspace{0.2cm}

}
\vspace{0.3cm}
\end{table*}

The soft component of the spectrum is modeled by blackbody disc
emission {\mtt with $l_s$ related to the power
supplied in the form of soft seed photons}, while the hard tail is
attributed to the corona, where photons from the disc repeatedly Compton
scatter off electrons with a hybrid thermal/non-thermal distribution.
\mt{Electron contributions are {\mtt then} parametrized by the
compactness parameters for thermal ($l_{th}$) and non-thermal
($l_{nth}$) electrons, {\mtt and we can define a compactness
parameter for the total power supplied to the electrons}, $l_h =
l_{th} + l_{nth}$.}
Typically, the corona non-thermal compactness has
{\mtt a comparable value} in both hard and soft Cyg X-1 spectral
states ($l_{nth} \sim 5$; Malzac et al., 2010); on the contrary,
most of the difference between the two spectral states is expected
to be due to a change in the soft photon compactness, $l_{s}$ (Malzac et al., 2010).

{\sasasa 
 For our analysis {\mtav of the soft state},
 we {\mtav considered  a} class of
hybrid Comptonization models,
and {\mtav fitted} the available data with EQPAIR, exploring how
the {\mtav relevant physical parameters (most importantly, the
soft compactness $l_s$ and the non-thermal to thermal compactness
ratio $l_{h}/l_{s}$), 
affect  the spectral energy distribution.} {\mtav
Our first goal is to determine a model with `extreme' parameters
that lead to a high energy emission {\it just} consistent with our upper limit above
100 \mev . {\sasasasa In all fits  a power-law distribution of accelerated/IC-cooled
electron/positron pairs is assumed {\mtt ($ n_{inj}(\gamma)
\propto \gamma^{-(\Gamma_{inj}+1)}$) with an injection index
$\Gamma_{inj} \sim 2.7$ and minimum and maximum electron/positron
Lorentz factors $\gamma_{min}$ and $\gamma_{max}$
}fixed to the values of 
1.3 and $10^3$ respectively, according to the well established literature
(Gierlisnki et al., 1999; Frontera et al., 2001; Del Santo et al., 2013 and ref. therein).
The non-thermal to total hard
compactness ratio $l_{nth}/l_{h}$ is set of order of unity in order 
to maximize the non-thermal component.
We have explored varying values of $l_s$ in the range 1-10, letting
$kT_s$, $l_{h}/l_{s}$, $\tau_i$ and $\Omega$ as free parameters.
This analysis in general
produces spectra incompatible with the whole set of data {for $l_s
< 10$}, predicting a persistent high energy component incompatible with
{\it AGILE} upper limit.
Our constraints to the parameter
space lead to a lower limit for the soft compactness, that is
constrained to be in the range $l_s
\gtrsim 10$ in order to be simultaneously consistent with both {\it RXTE} data 
and {\it AGILE}
upper limit, given the adopted value for
$\gamma_{max}$\footnote{
Note that for a value of the injection index of $\sim 2.7$, higher values of $\gamma_{max}$
would have negligible effects on the results, since only a small power is injected at this
energy. The maximum allowed value of $\gamma_{max}=10^4$ is however not completely 
consistent with {\it AGILE} upper limit, producing some power around 100 \mev .
 }.
We therefore proceeded by freezing the soft inner disk component to $l_s = 10$
in order to determine the parameters reported in table 1.  
We show in Fig. \ref{fig:gierlinski02} the spectral energy distributions
and in Tab. 1 the results of the fitting procedure for the two data sets.
{\mtav 
{{\it AGILE} upper
limit obtained during the soft state is in red}. Superimposed to
the models are the {\it RXTE} PCA/HEXTE data after the spectral
transition (green-colored data are for the model-1 soft state of 4$^{th}$
of July,  and cyan-colored data are for the model-2 super-soft state of
July 22$^{nd}$, 2010). We also show, for comparison, in black
color, the historical COMPTEL gamma-ray data points for the
Cyg X-1 soft state detection\footnote{\mtt Note that this detection
constitutes a single (and so far unique) episode of emission above
1 \mev , and that another observation by COMPTEL in the
soft state during January 1994 did not detect any emission from
Cyg X-1.} during June 1996, and the model by McConnell et al. (2002) 
for these data with a black dashed line.

{\mtav 
We notice that both `extreme' models tend to
predict higher gamma-ray fluxes in the range 1--30 \mev\ than what measured
in the historical COMPTEL detection. We notice however that a more realistic modelling would 
require more broad band data
to better constrain the values for $l_s$,
$l_{nth}/l_{h}$ and $\Gamma_{inj}$.}

%

}

{\mtav Our model-1 is in qualitative agreement with model parameters
explored in Gierlinski et al. 1999 for the soft state.} We add the
crucial information of the non-existence of a strong non-thermal
component of accelerated electrons/positrons with a power-law
index harder than $\Gamma_{inj} = 2.7$.} 
The ratio of $l_{h}$/$l_{s}$ is well constrained to values
$< 1$, as for typical soft states. From the constraints to the soft
compactness we can therefore extrapolate a range of possible
values for the hard compactness (and consequently the non-thermal
and thermal compactness), \mt{ obtaining $l_{h} \gtrsim 6$.}}

\section{Conclusions}

{\mtt The prolonged soft state of Cyg X-1 in mid-2010/mid-2011
offered an unprecedented opportunity to verify the existence of a
prominent non-thermal tail in the gamma-ray spectrum of a black
hole system in accretion above 10 \mev\ (i.e. COMPTEL data).
Our {\it AGILE} observations exclude the
existence of such a tail. This result, combined
with previous observations of Cyg X-1, confirms the
physical picture of this state based on soft thermal X-ray
emission emanating from the inner disk and partial reprocessing
and scattering by a corona. {\sasasa It is interesting to note that whereas
the ratio parameters $l_{h}$/$l_{s}$ and $l_{nth}/l_{h}$ are
similar to previous Cyg X-1 soft states detected 1994 and
1996 (e.g., Gierlinski et al. 1999), we find a quite
well constrained value for the compactness, related to feeding soft
seed photon luminosity $l_s \gtrapprox 10$. }
We believe that our measurements, exploring and combining data in energy ranges
much broader than in past analyses, constitute the most accurate
constraints on the underlying physical processes thus far.}

{\mtt By considering both hard and soft state upper limits to the
emission from Cyg X-1, we can put our measurements in perspective.
Cyg X-1 spends most of its time in a sub-Eddington optically thick
hard state. Occasionally, the accreting system dramatically
changes its configuration to the soft state. The overall (mostly
soft X-ray) luminosity increases by {\sasasa a factor up to 3 in magnitude} (Zdziarski et al., 2002)
getting closer to the Eddington luminosity. In this state, the coronal processes can be
revealed more easily because of the optical thinness of the
corona. We find that there are no major variations, on the
average, of the conditions that lead to the energization of a
non-thermal population of electrons/positrons compared to the hard
state. The average emission properties of Cyg X-1 at energies above
1--10 \mev\ appear to be quite stable.}

{\mtt  We notice that this behavior of Cyg X-1 is in contrast with
even the average properties of the other prominent Galactic
micro-quasar Cygnus X-3 (Tavani et al., 2009; Abdo et al., 2009). {\sasasa In
the case of Cygnus X-3, gamma-ray emission above 100 \mev\ is
clearly transient and originates in states with a relatively low hard X-ray flux.}
With the exception of two episodes of transient gamma-ray emission detected
by {\it AGILE} from Cyg X-1 and reported in the Appendix, such an activity
of recurrent and very active transient emission is not the norm in
Cyg X-1.}

{\mtt Transient gamma-ray emission from Cyg X-1 originating from physical
processes different from those of a `steady' disk$+$corona can be
difficult to detect. The very short (less than 2
hours) TeV emission detected by MAGIC from Cyg X-1, if confirmed, is
quite remarkable. The current gamma-ray missions {\it AGILE} and {\it Fermi}
can detect gamma-ray variability at the level of hours only for
very intense events. In the Appendix, we report one of these
candidate transient events from Cyg X-1 , which was detected by {\it AGILE}
during the transition from hard to soft state on June 30$^{th}$ to July 2$^{nd}$
2010. If confirmed, this class of transient gamma-ray emission would
open a new window into the physical processes around
accreting black holes, allowing the possibility of jet or `pre-jet'
launching activity of these transient events.
Cyg X-1 transient gamma-ray activity could occur on
short timescales (of order of the day or shorter)
and with a typical gamma-ray flux of $F_{\rm \gamma}\sim 100-150 \times 10^{-8}$ \flux.
Such events would be difficult to be detected by the current generation of gamma-ray
telescopes ({\it AGILE, Fermi}). Future instruments with an improved exposure will allow us to
investigate these issues in much more detail.
}

{\mtt }

\section*{Acknowledgements}
{\sasasa We thank the anonymous referee for his/her careful reading and for the 
important suggestions that considerably improved the quality
of the manuscript. Research partially supported by the ASI grants nos. I/042/10/0 and I/028/12/0.
MDS acknowledges financial support from the agreement
ASI-INAF I/009/10/0 and from PRIN-INAF 2009 (PI: L. Sidoli).
}

\appendix
\section{A review of gamma-ray observations of Cygnus X-1 above 1 {\bf \mev}}

{\mta We summarize in this Appendix all relevant observations and
possible detections of Cyg X-1 above 1 \mev . We briefly describe the
(so far) unique high-significance COMPTEL detection of Cyg X-1 up to
5--10 \mev\ in June 1996. A short (less than 2 hours) episode of
emission at TeV energies was detected by MAGIC in
2007. Finally, we discuss the gamma-ray event above 100 \mev\
detected by {\it AGILE} in pointing mode in October 2009 (Sabatini et
al., 2012a), and focus on a new possible event detected by {\it AGILE}
in spinning mode in early July 2010 in coincidence with a dramatic
spectral change from hard to soft states.}

\subsection{Gamma-ray observations of Cygnus X-1 in the Soft State in 1994 and 1996:  \bf COMPTEL data.}
\label{sec:softstatedata}

Observations of Cyg X-1 during the soft state in the gamma-rays
are {\mta scarce} in the literature, also due to its intrinsic
behavior: the source spent 90$\%$ out of its time in the hard
state during the last $\sim 20$ years. During the operational
period of {\mta the Compton Gamma-Ray Observatory, {\it CGRO}
(1991-2000) the instruments onboard (BATSE, OSSE, COMPTEL, EGRET)
observed several times the Cygnus region.} Cyg X-1 was in a
clear soft state in only two occasions: in January 1994 and in May
1996. In both cases, {\it CGRO} pointed at the source with a ToO
following the announcement of the hard-to-soft state transition.
For the 1994 event (VP 318.1) all four {\it CGRO} instruments collected
data, while for the 1996 one (VP 522.5) EGRET was switched off.
Fig. \ref{fig:batse1994} shows the BATSE long term lightcurve for
the 1994 soft state and the {\it CGRO} ToO time period (marked by vertical dashed lines).
No simultaneous soft X-ray monitoring was available
at that time. COMPTEL did not detect any emission from Cyg X-1
for this period, and the upper limit was consistent with the
$E^{-2.7}$ power law measured by both BATSE (Ling et al., 1997)
and OSSE (Phlips et al., 1996).

\begin{figure}[h!]
\begin{center}
\includegraphics [width=6. cm, angle=90]{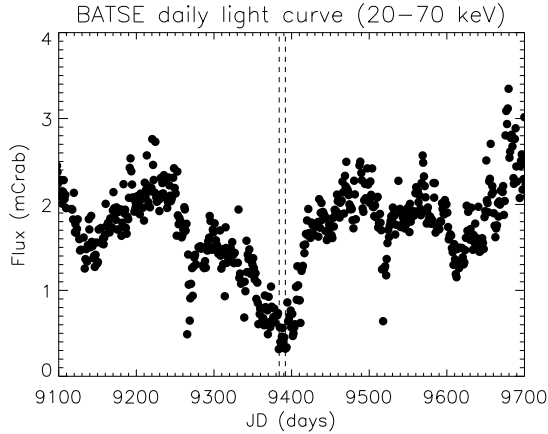}
\caption{\mt{ The Soft Spectral State of Cyg X-1 in January 1994: BATSE light curve
and COMPTEL observing period in dashed lines (VP 318.1, January 1994). No emission was
{\mtt detected by COMPTEL or EGRET from Cyg X-1 above 1-10 \mev}
during this period. }}\label{fig:batse1994}
\end{center}
\end{figure}

\begin{figure}[h!]
\begin{center}

\includegraphics [width=6. cm, angle=90]{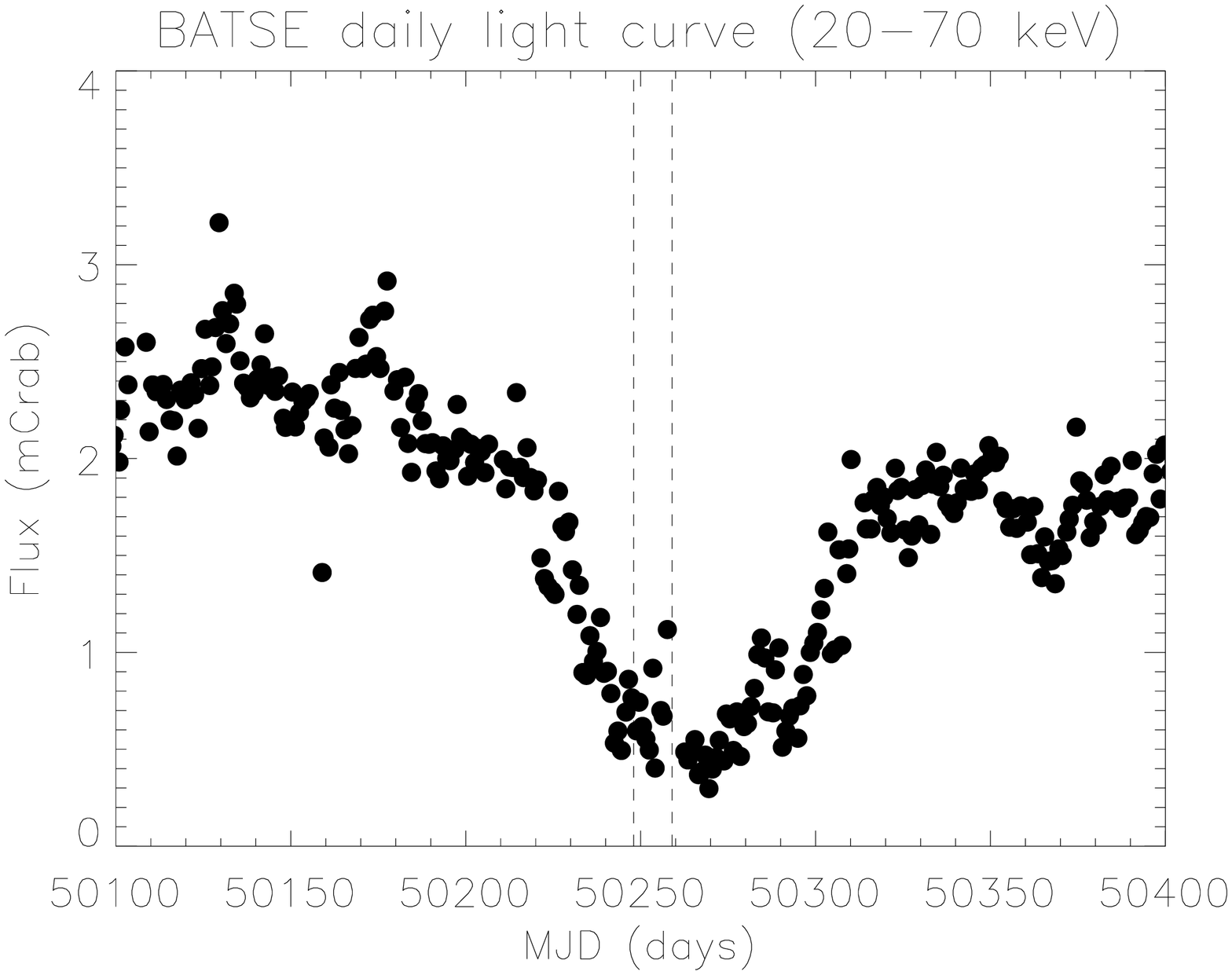}
\includegraphics [width=6. cm, angle=90]{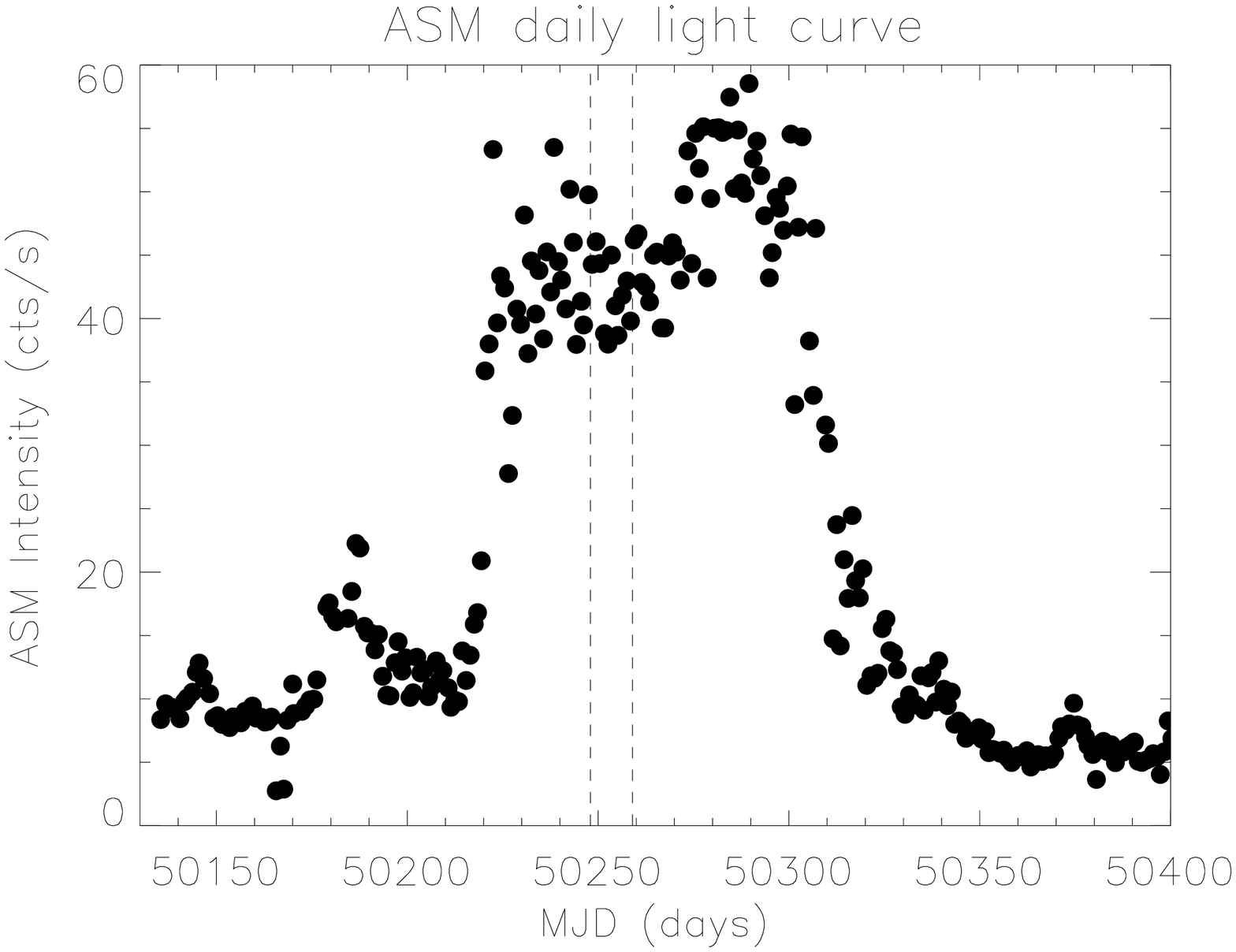}

\caption{\mt The Soft Spectral State of June 1996:  BATSE (left panel) and ASM A-band (right panel) light
curves and COMPTEL observing period in dashed lines (VP 522.5).
COMPTEL has detected Cyg X-1 in the range 1-10 \mev for
this period (Mc Connell et al. 2002). }\label{fig:batse1996}
\end{center}
\end{figure}





{\mta Another interesting soft state episode occurred in June-July
1996}. Fig. \ref{fig:batse1996} shows the BATSE and simultaneous
ASM long-term data around the 1996 Cyg X-1 soft state; the  CGRO ToO
viewing period {\mta is marked with vertical dashed lines}. This
observation, with a more favourable angle in the Field of View,
 {\mta resulted in } the first gamma-ray detection
above $\sim 1$ \mev of Cyg X-1. The {\mta hard X-ray} spectral
index was similar to that of the 1994 event ($\sim -2.5$).
The overall intensity was also measured by OSSE to be higher than before of
about a factor 2 {\mta (McConnell et al., 2002)}. This particular episode
has been considered the 'canonical' soft spectral state for a long time. 
The expectation from the model is that part of the emission
should also appear at energies $\ge 100$ \mev , while {\it AGILE} shows that no emission
is detected in this energy range, with an upper limit of
0.01 \kev\ $\rm cm^{-2}$ $\rm s^{-1}$ (see Fig. \ref{fig:gierlinski02}).

{\mt \subsection{Transient gamma-ray episode of Cyg X-1 in the hard
state:  MAGIC observations}}

{\mt The Cyg X-1 hard state is described by a power law of typical
spectral index 1.7 {\mta in the hard X-ray range}, and a sharp
energy cut-off around 150 \kev. Therefore, significant gamma-ray
emission is not expected in this spectral state. Until recently
the higher energy data available in the literature were those of
COMPTEL (McConnell et al., 2000; McConnell et al., 2002), in agreement
with this picture.
EGRET provided only an upper limit
for the source in the hard state (Hartman et al., 1999).}

{\sasa MAGIC reported for the first time an episode of transient TeV emission
from Cyg X-1 in 2007 (Albert et al., 2007).}
The spectral state during this observation was a typical hard state
spectrum and no unusual feature in the X-ray light curve and
spectrum was noted. {\sasa Quasi-simultaneous observations were
carried out by {\it INTEGRAL}: the TeV detection coincides with the peak of a
small X-ray flare just after a very fast rise in hard X-ray flux, but
no obvious correlation between the X-ray and TeV emission was found (Malzac et al., 2008).}

\sasa{\subsection{Transient gamma-ray episode of Cygnus X-1 in the hard
state: {\it \bf AGILE} observations}}

{\mta As reported in Sabatini et al 2010a, also {\it AGILE} detected
above 100 \mev\ a fast ($\sim 1\, $day) transient event from Cyg X-1 in October
2009 during a hard state period. Although not simultaneous with the
MAGIC event, the {\it AGILE}
detection of a gamma-ray flare during a hard state,
of the duration of the day or shorter, further suggests that
{\sasasa additional non-thermal} components {\mta may} appear also in states
previously believed to be characterized by a cut-off above a few
\mev . The {\it AGILE} map {\mta of the October 2009 gamma-ray event} is shown in
Fig \ref{fig:flare_hardstate}.
Here we also shows the multi-wavelength ( AMI-LA,
MAXI and {\it Swift}-BAT) daily monitoring of Cyg X-1 during the
gamma-ray flare detected by {\it AGILE}: as for the MAGIC flare, there
is no evidence of {\mta detectable} spectral changes or unusual
features on the {\it day timescale.}} {\sasasa It is however interesting to point out
that a blind search analysis carried out in about 4 years of {\it Fermi} data shows that
some low significance activity is present in the gamma-ray data
above 100 \mev\ during the periods of this gamma-ray flare (and the one discussed 
in sec. A.4.1) reported by the {\it AGILE} Team for Cyg X-1. The analysis was supported by 
a statistical treatment of spurious detections and other periods of gamma-ray activity
outside this ones and the one in sec A.4.1 reported by {\it AGILE} are 
probably spurious (Bodaghee, private communication; 
see also Bodaghee, 2012).}




\begin{figure*} 
\begin{center}
\includegraphics [width=10 cm]{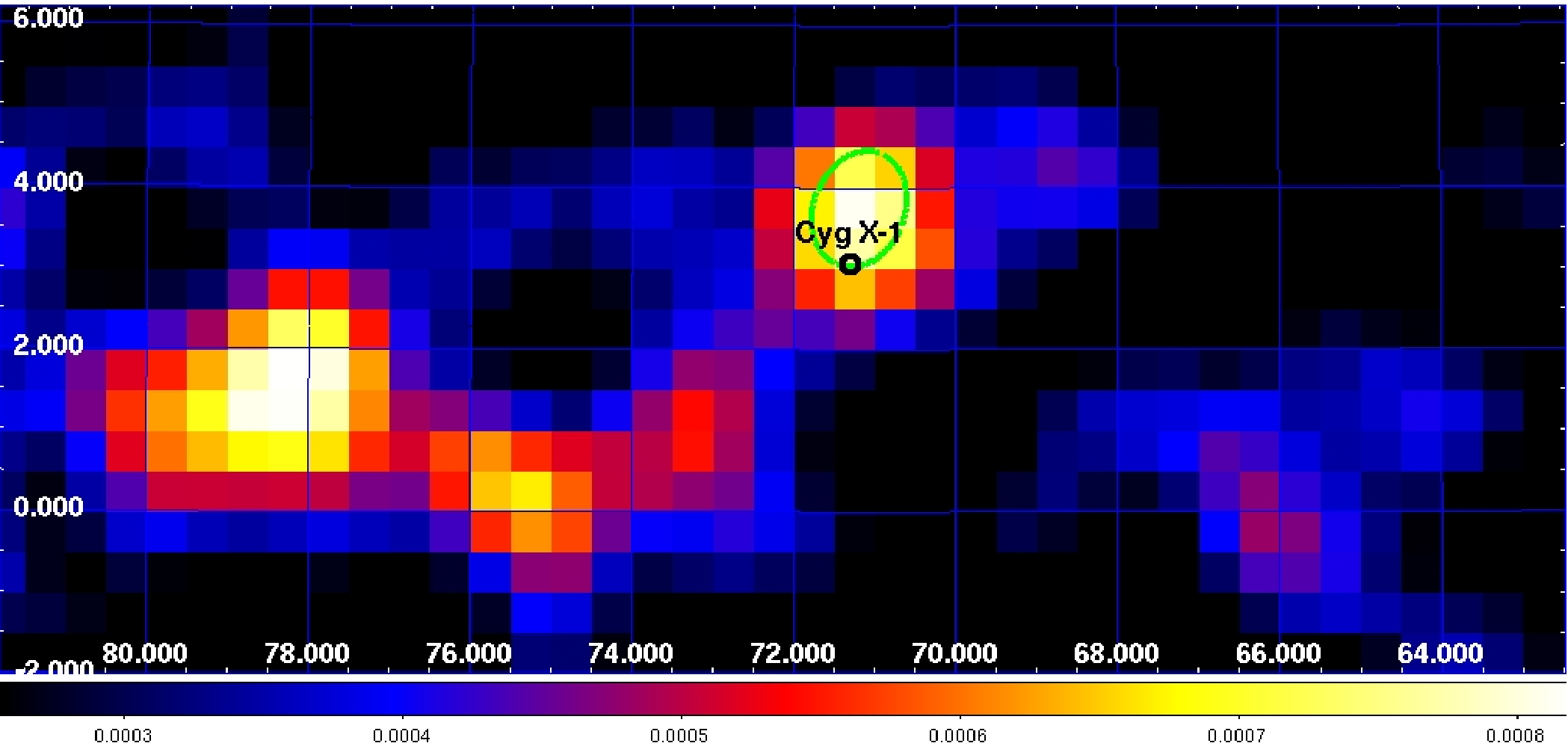}
\includegraphics [height=15 cm] {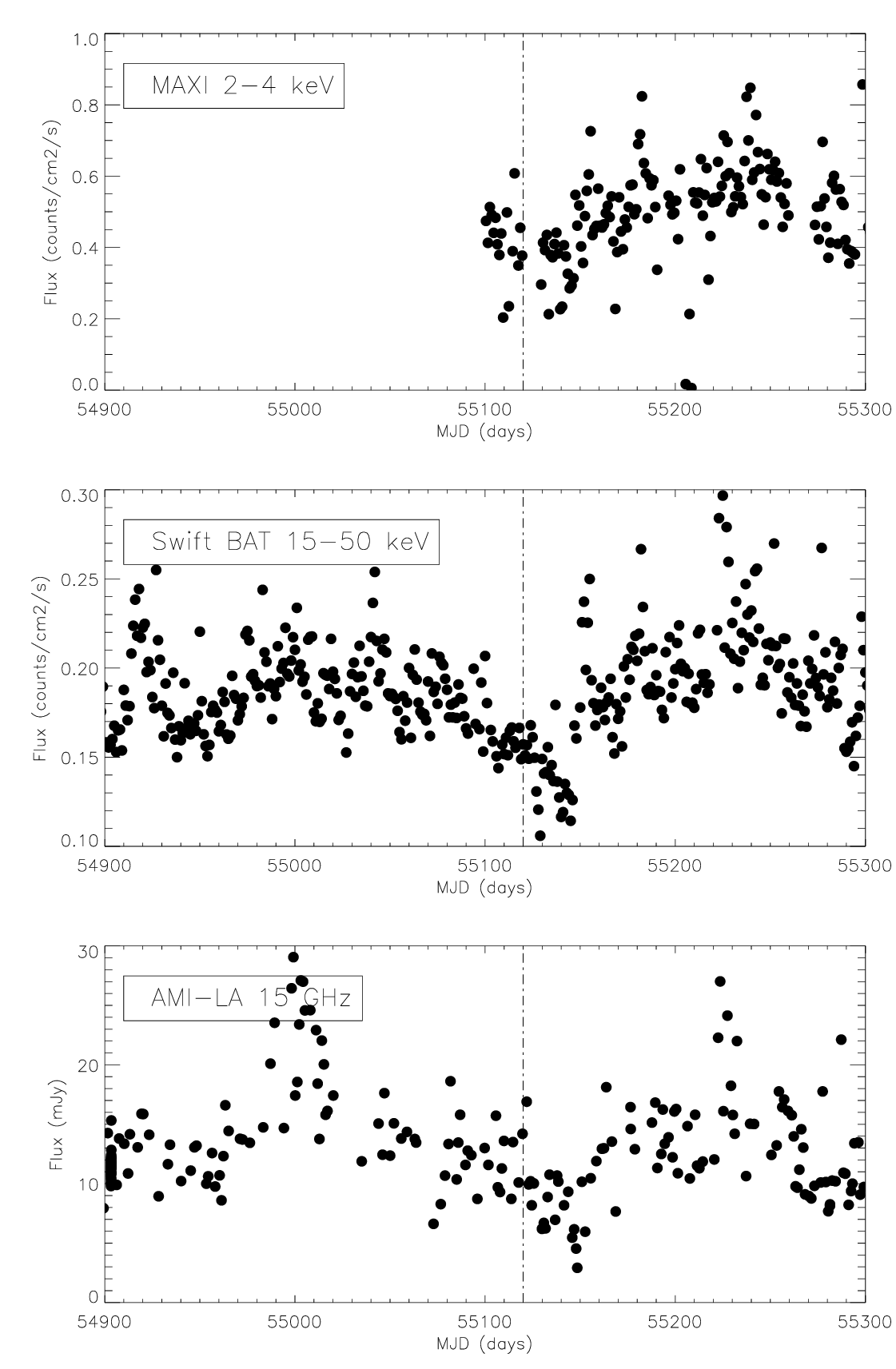}
\caption{\mt Hard Spectral State.{\it Upper panel:} {\it AGILE}
gamma-ray intensity map above 100 \mev\ of the Cygnus Region in
Galactic Coordinates displayed with a three-bin Gaussian smoothing
and a pixel size of 0.5$^\circ$. The map is obtained by
integrating data in the period 2009-10-15 UTC 23:13:36 to
2009-10-16 UTC 23:02:24. The black circle is the optical position
of Cyg X-1 and the green contour is the {\it AGILE} 2$\sigma$ confidence
level. Other panels show multi-wavelength daily monitoring of Cyg
X-1: {\sasa {\it Swift}-BAT data in the 5--50 \kev\ in the {\it upper panel};
{\it MAXI} data in the 2--4 \kev\ in {\it middle panel} and AMI-LA data at 15 \ghz\
in {\it lower panel}}. The {\mta vertical dashed lines show the
duration of the gamma-ray event reported in Sabatini et al.,
2010a.}} \label{fig:flare_hardstate}
\end{center}
\end{figure*}





\mt{ \subsection{The hard-to-soft state transition of June-July 2010: RXTE PCA data and {\it \bf AGILE} observations}}

After having spent \mt{a long period from} 2006 to mid 2010 in an
extraordinary hard state (Nowak et al., 2011), on the 28th of June
2010 Cyg X-1 entered in a transitional state, passing from the
hard to the soft state.  A gradual spectral softening of the black
hole during the period 10th of June -
 1st of July 2010 was announced by \mt{{\it MAXI/}GSC} \mt{(Negoro et al., 2010)} and the
subsequent soft X-ray\mt{ increasing} emission was also reported
by {\it RXTE/}ASM (Rushton et al., 2010a), confirming the
transition of the source from the hard to the soft spectral state.
The rapid fall in hard X-rays around June 29 - July 01 2010 was
also reported by {\it Fermi}-GBM (Wilson-Hodge et al., 2010).
A multi-wavelength campaign was triggered by the transition
episode, providing a {\mta wealth} of data from gamma-rays to
radio ({\it MAXI}, Negoro et al., 2010; {\it RXTE/}ASM, Rushton et al., 2010a;
{\it AGILE}, Sabatini et al. 2010b; {\it Fermi}-GBM, Wilson-Hodge et al.,
2010; {\it SWIFT}, Evangelista et al., 2010; MERLIN, Rushton et al.,
2010b; WRST, Tudose et al., 2010). All observations showed the
source to be in a intermediate-soft state \mt{(Belloni et al,
1996)}. The source was detected to be in the soft state on the
11th of July 2010 (Rushton et al., 2010b), and remained
in this state until the end of April 2011 (Grinberg et
al., 2011). \mt{ Fig. \ref{fig:gpooley}} shows a multi-wavelength
long-term monitoring of the 2010-2011 soft state {\mta  in the
hard X-rays (BAT 15-50 \kev),  soft X-rays ({\it MAXI} 2-4 \kev) and radio
(AMI-LA 15 \ghz)}. The {\mta vertical} dotted-dashed lines show
the {\mta duration of a candidate episode of enhanced gamma-ray
emission} detected by {\it AGILE} during {\mta the remarkable
hard-to-soft transition of July 2010}.

As reported in the main text, nineteen pointed observations were performed by {\it RXTE}-PCA during
the period 2010 June 19 - 2010 July 31, for a net exposure time of
about 68.5 ks, catching the source across the whole transition from the
hard to the soft state.  The observations were carried out in the binned data mode
(B-2ms-8B-0-35-Q), with 1.95 ms bin size in the energy band
2.1-14.8 \kev. In Fig. \ref{fig:PCA} we plotted the {\mta X-ray power spectrum}
(normalized to units of fractional squared RMS) of the {\it RXTE}-PCA
observation 95121-01-13-00 (2461 s net exposure) carried out on
2010-06-19 with T$_{start} = 21:44:26.3$~UT \mt{(black line)},
for the observation 95121-01-14-00 (1730~s net exposure) performed
on 2010-07-04 with T$_{start} = 03:27:02.6$~UT  \mt{(red line)} and
for the observation 95121-01-13-00 of the 2010-07-22 with  T$_{start} = 07:40:40.28$~UT.
The {\it RXTE}-PCA data clearly show a variation in the noise components
of the power spectra ({\it PDS}), with a decrease in the RMS variability
during the state change. The fractional RMS was $\sim$8\% on the 19${th}$
of June 2010, with a power spectrum
showing band-limited noise between 0.3 \hz\ and 10 \hz\ (Fig.
\ref{fig:PCA}, \mt{grey line}) consistent with an intermediate state
(see, e.g., Shaposhnikov \& Titarchuk, 2006). The fractional RMS
then dropped to about 4\% on 2010 July 4, with a narrower noise
component in the PSD which peaks at $\sim$3 \hz\ (Fig\ref{fig:PCA}, left panel,
\mt{green line}), thus showing that the source \mt{had} finally
reached the soft state. {\sasasa We also plot in cyan the PDS of the 22$^{nd}$ of July, clearly
showing a super-soft state, as an example of the intrinsic variability present in the
soft state period  monitored by {\it AGILE}.}
Although not simultaneous with the {\it AGILE} candidate flaring
event (see next section), these observations are of particular interest to
the gamma-ray data because they are few days before and just
after the gamma-ray possible detection, suggesting the coupling
 of transitional states with gamma-ray emission.

\begin{figure*} 
\begin{center}

\includegraphics [height=20 cm, angle=0] {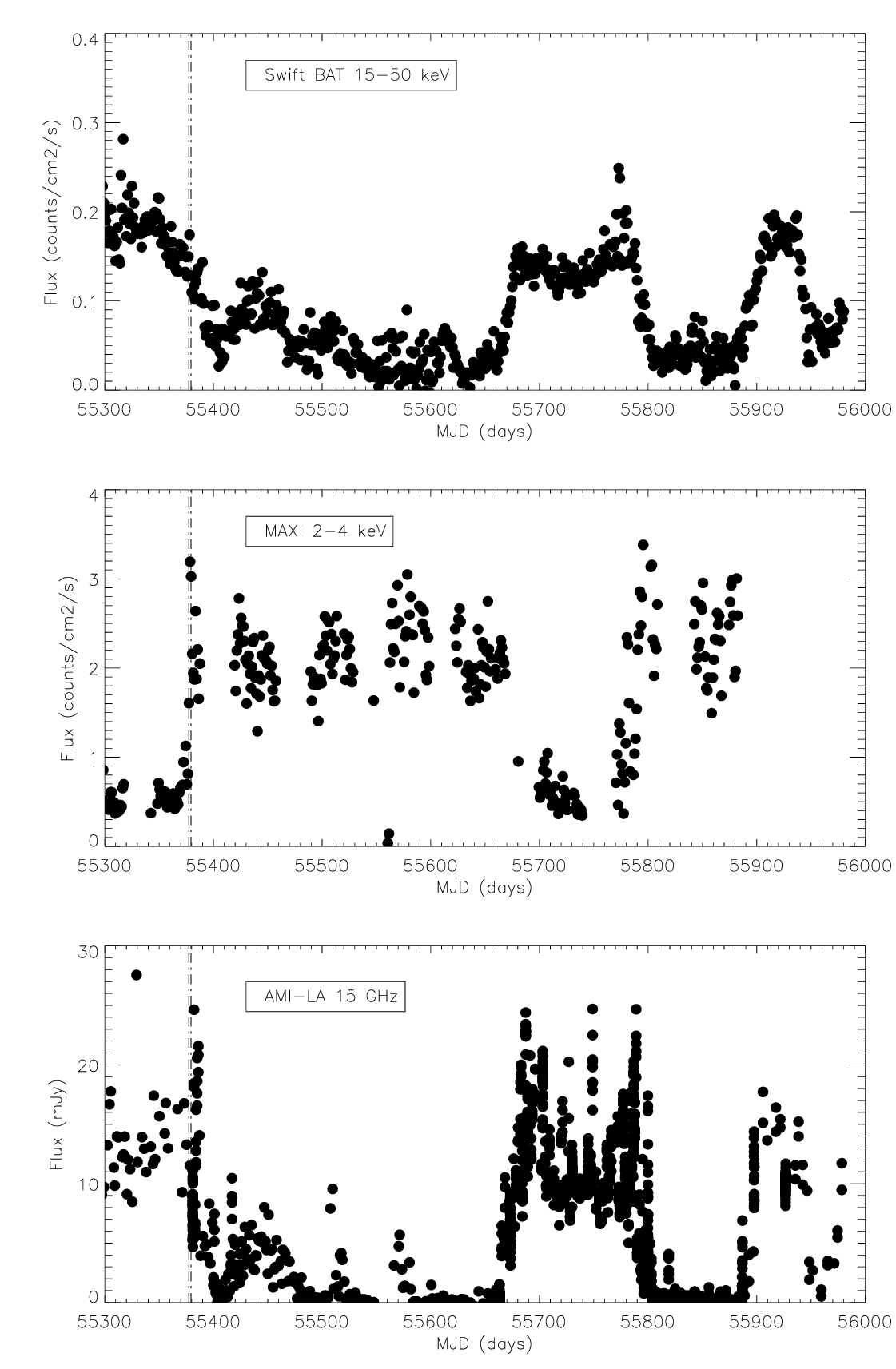}
\caption{\mt Multi-Wavelength daily monitoring of Cyg X-1. Upper panel
showes {\it Swift}-BAT data in the
15-50 \kev\ energy range, middle panel  {\it MAXI} data in the 2-4 \kev\ band
  and lower panel AMI-LA data at 15 \ghz. Dashed lines
refer to {\it AGILE} candidate flaring event.}
\label{fig:gpooley}
\end{center}
\end{figure*}

\subsubsection{An {\it AGILE} possible detection of Cygnus X-1 at the hard-to-soft transition in July 2010}

\begin{figure} 
\begin{center}

\includegraphics [height= 8 cm]{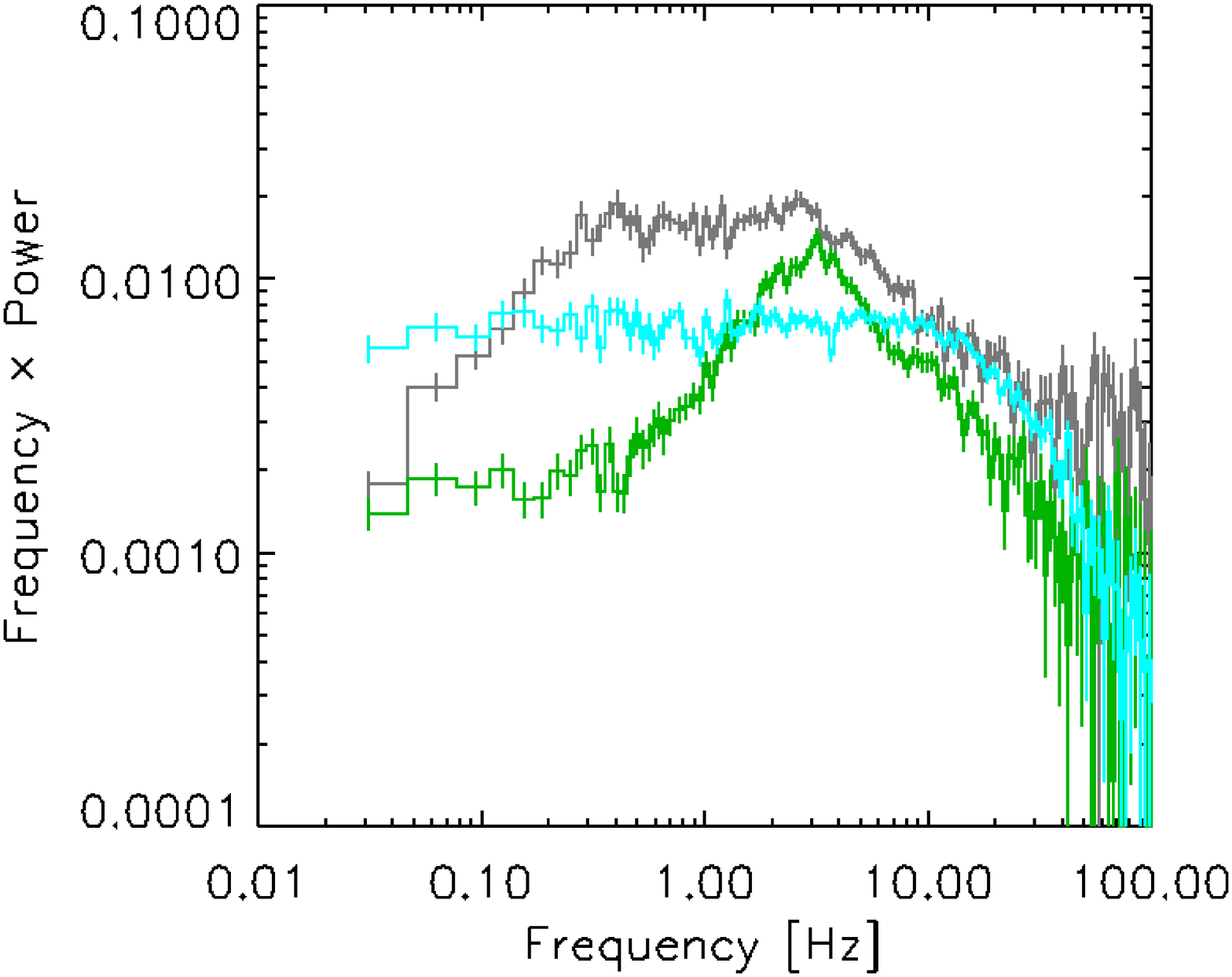}
\includegraphics [height=8 cm]{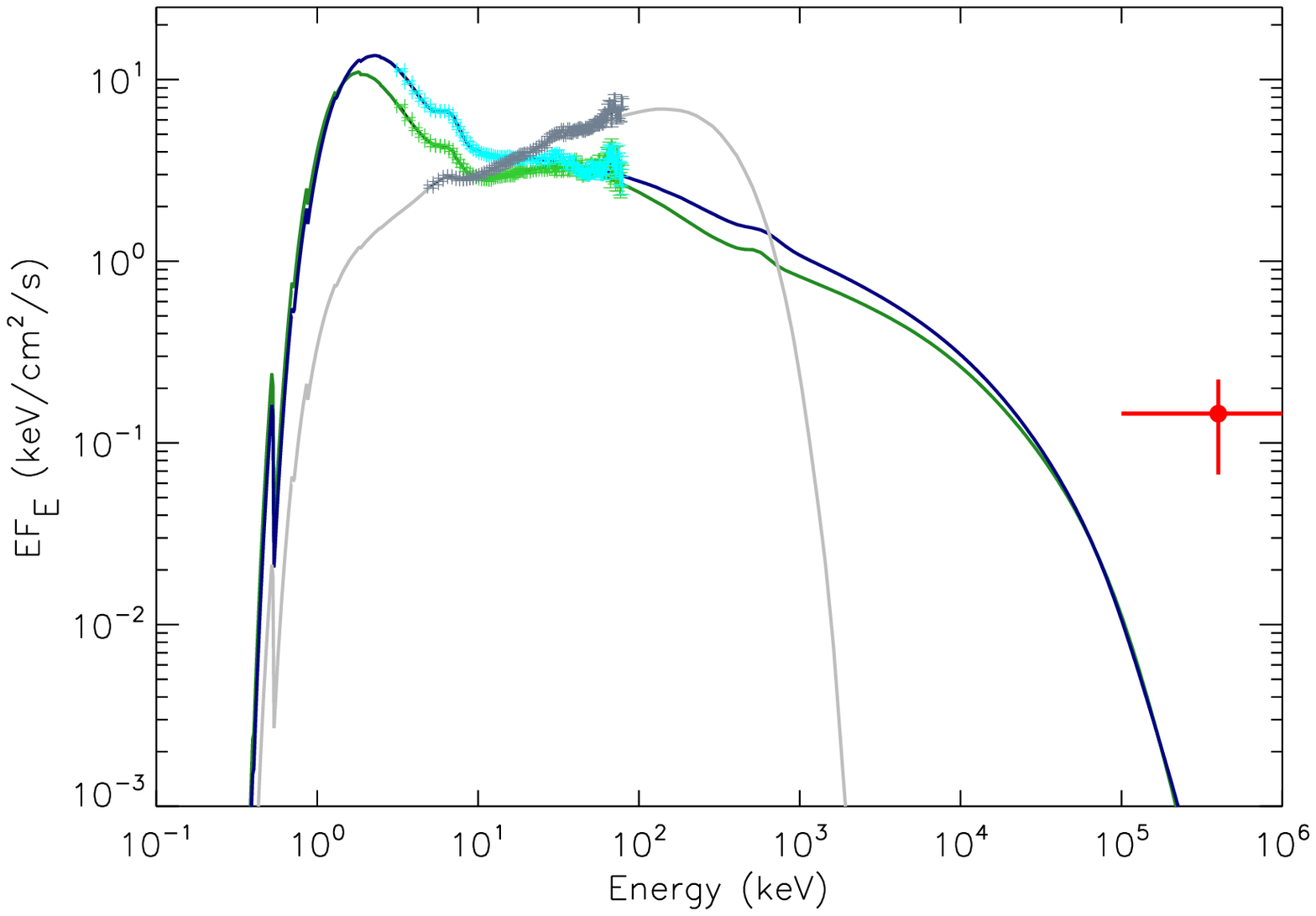}


\caption{\mt {\it Left panel:} Power Density Spectra of Cyg X-1 before and after 
the spectral transition
occurred at the end of June 2010. RXTE PCA ToO data on the 2010-06-19
 is the grey curve; 2010-07-04  is the green curve and 2010-07-22 is the cyan curve respectively.
{\it Right panel:} Corresponding Spectral Energy Distribution
with {\it RXTE} PCA/HEXTE data for the three days as in left panel and {\it AGILE} flare in red.}
\label{fig:PCA}
\end{center}
\end{figure}

\begin{figure} 
\begin{center}

\includegraphics [height=5 cm]{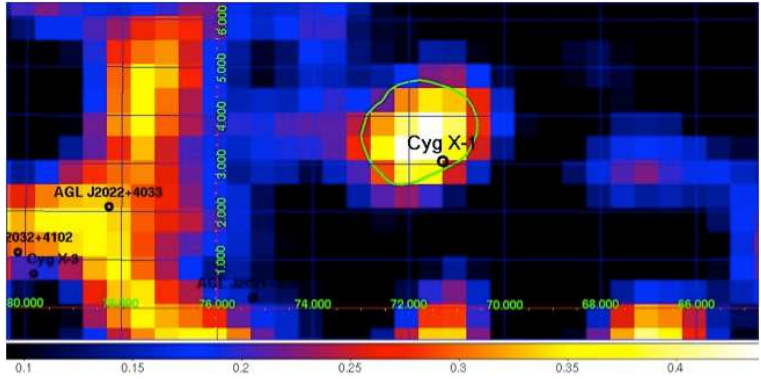}


\caption{\mt
{\it AGILE} candidate transient event on June 30$^{th}$ -- July 2$^{nd}$ 2010.
Gamma-ray intensity map above 100 \mev\ of the Cygnus
Region in Galactic coordinates displayed with a three-bin Gaussian smoothing and a
pixel size of 0.5$^\circ$. The map is obtained by integrating data in the period: 2010-06-30
10:00 UT to 2010-07-02
10:00 UT. The nominal position of Cyg X-1 is overlayed in back and the error box
of the detection is in green. The color bar scale is in units of
photons cm$^{-2}$s$^{-1}$.}
\label{agilemap2}
\end{center}
\end{figure}

\begin{figure} 
\begin{center}

\includegraphics [height=15 cm, angle=0] {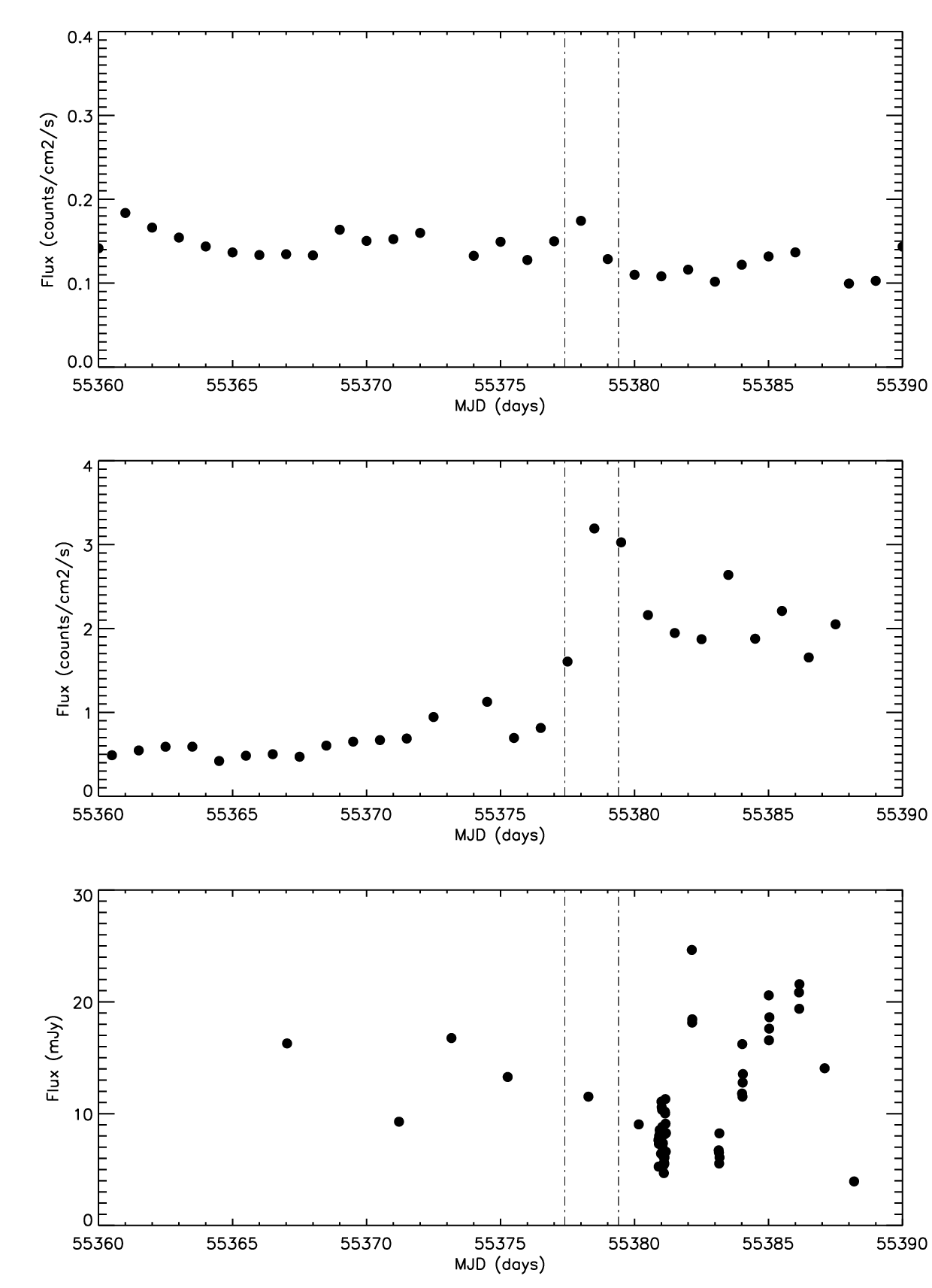}

\caption{\mt Multi-wavelength daily monitoring of Cyg X-1 focussing on the hard-to-soft transition
of June 2010. Upper panel
shows {\it Swift}-BAT data in the
15--50 energy range, middle panel  {\it MAXI} data in the 2--4 \kev band
  and lower panel AMI-LA data at 15 \ghz. Dashed vertical lines
refer to the {\it AGILE} candidate flaring event on June 30$^{th}$ -- July 2$^{nd}$ 2010.
} \label{fig:guypooleyzoom}
\end{center}
\end{figure}


 \mt{We carried out an automatic search for transient
gamma-ray emission in {\it AGILE} data during the whole {\mta 2010--2011
period}, and found evidence of gamma-ray activity during the 2010
hard-to-soft spectral transition. {\mta Based on previous claims
of gamma-ray detections of Cyg X-1 on short timescales by MAGIC
(Albert et al., 2007) and {\it AGILE} (Sabatini et al., 2010a), we searched
 for events occurring {\sasa on short time
scale (2-days). A relatively weak, i.e. low statistical significance,
 but interesting gamma-ray enhancement occurs exactly at the hard-to-soft
transition at the end of June 2010.}} Integrating from 2010-06-30 10:00 UT
to 2010-07-02 10:00 UT, the maximum likelihood analysis
yields} a flux excess above 100 \mev\
of $F_{\rm \gamma}= 145 \pm 78 \times 10^{-8}$ \flux  with a 3$\sigma$ 
statistical significance.
\mt{Fig. \ref{agilemap2}} shows the {\it AGILE} gamma-ray intensity map of the
Cygnus region above 100 \mev\ for this period.
{\sasasa Although not simultaneous, we think it is interesting to 
show in Fig. \ref{fig:PCA} the {\it AGILE} data point for the 
candidate flare with the extreme models (model-1 and model-2) 
discussed in the main text. For comparison, we also show in grey 
the {\it RXTE} PCA/HEXTE data for the ToO observation of 
the 19th of June 2010, i.e. 10 days before the {\it AGILE} candidate flare, when Cyg X-1 was 
in a hard/intermediate state (we plot a representative model with 
$l_{nth}/l_{h}=0$ for this case).}

Although the statistical significance of the gamma-ray {\mta
enhancement detected by {\it AGILE}} is low {\mta (because of the poor
statistics obtainable for short events by {\it AGILE} in spinning
mode)}, it is interesting to discuss this {\mta candidate} event in a
multi-wavelength perspective.
Fig. \ref{fig:gpooley} shows a long-term monitoring in
hard X-rays ({\it Swift}-BAT, {\it upper panel}), soft X-rays ({\it MAXI} in
the 2--4 \kev\ band, {\it middle panel} ) and radio (AMI-LA 15 \ghz\
band, {\it lower panel}); the dashed lines show the {\it AGILE}
detection. Interestingly, the gamma-ray flare happens to be
simultaneous with the definitive transition to the soft state, and
anticipates by about 2 days {\mta an `anomalous' intense radio
flare detected well in the soft state (Rushton et al., 2012), occurring
therefore when shocks are possibly predicted to be formed within the jet
(Fender et al., 2004).  {\sasasa As already mentioned in sec. A.3, a 
blind search analysis supported by a statistical treatment of spurious detections 
shows that some low significance activity is present also in the {\it Fermi} gamma-ray 
data during the period of this gamma-ray flare (Bodaghee, private communication; 
see also Bodaghee, 2012).}

Fig. \ref{fig:guypooleyzoom} shows the detailed transition as detected in the
hard X-rays (BAT), 2--4 \kev\ X-rays ({\it MAXI}), and radio (AMI-LA).
The time period of enhanced gamma-ray emission above 100 \mev\
possibly detected by {\it AGILE} is marked by vertical dashed lines}.

{\mta We also searched for gamma-ray activity from Cyg X-1 in
coincidence with other interesting spectral transitions as shown in
Fig. \ref{fig:gpooley}. However there is  no evidence of
enhanced emission in the data ($F_{\rm UL} \sim 200 \times 10^{-8}$ \flux
for 2-days integration).
Fig. \ref{fig:gpooleyzoom2} shows the detail of the other recent
hard-to-soft transition which occurred in January 2011 and
led to another prolonged soft state ($\sim$ MJD: 55800 -- 55890). We
note that in this case the hard-to-soft transition occurs on a
timescale of several days, i.e., much longer than the sharp
transition recorded in July 2010 in coincidence with the {\it AGILE}
candidate event.}


\begin{figure} 
\begin{center}

\includegraphics [height=15 cm, angle=0] {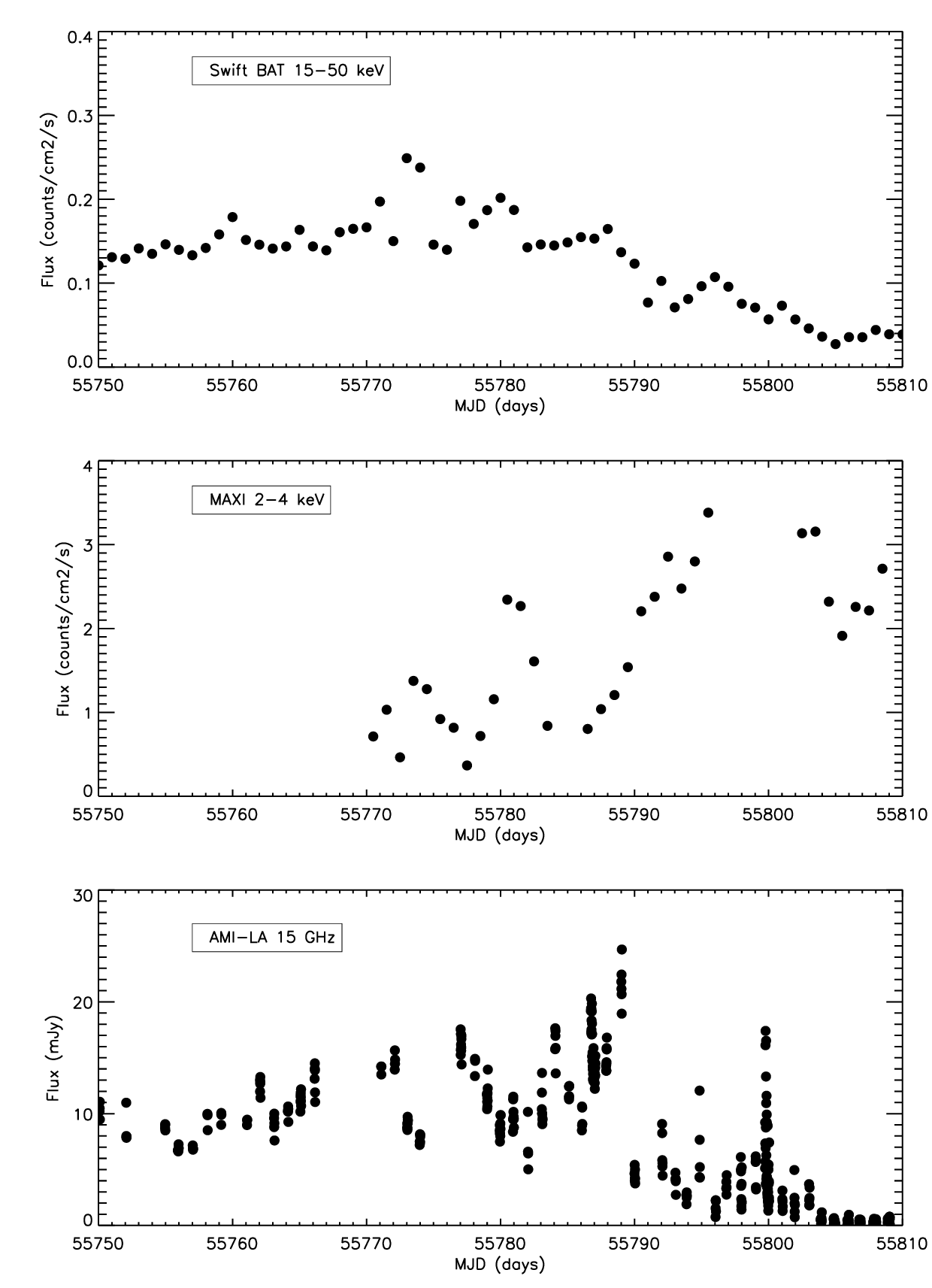}

\caption{\mt Multi-Wavelength daily monitoring of Cyg X-1 focussing on the hard-to-soft transition
of January 2011. Upper panel
showes {\it Swift}-BAT data in the
15--50 \kev\ energy range, middle panel  {\it MAXI} data in the 2--4 \kev\ band
  and lower panel AMI-LA data at 15 \ghz.
} \label{fig:gpooleyzoom2}
\end{center}
\end{figure}

\end{document}